\documentclass[acmsmall]{acmart}
\AtBeginDocument{%
  \providecommand\BibTeX{{%
    \normalfont B\kern-0.5em{\scshape i\kern-0.25em b}\kern-0.8em\TeX}}}

\setcopyright{acmcopyright}
\copyrightyear{2021}
\acmYear{2021}
\acmDOI{10.1145/1122445.1122456}

\acmJournal{TOSEM}
\acmVolume{37}
\acmNumber{4}
\acmArticle{111}
\acmMonth{11}

\usepackage{microtype}
\usepackage{textcomp}
\usepackage{bm}
\usepackage[caption=false]{subfig}
\usepackage{flushend}
\usepackage{csquotes}
\usepackage{lscape} 
\usepackage{booktabs} 
\usepackage{framed}
\usepackage{graphicx}
\usepackage{tabularx}
\usepackage{longtable}
\usepackage{graphicx}
\usepackage{url}
\usepackage{multirow}
\usepackage{enumitem}
\usepackage{hyperref}
\usepackage{lscape}
\usepackage[normalem]{ulem}
\useunder{\uline}{\ul}{}
\usepackage{latexsym}
\usepackage{xcolor}
\definecolor{lightgray}{gray}{0.9}
 
\usepackage{siunitx}  
\usepackage{textgreek}  
\usepackage{wasysym}
\usepackage{makecell}

\usepackage{tabularx}

\begin{document}

\title{Is It Safe To Learn And Share? On Psychological Safety and Social Learning in (Agile) Communities of Practice}


\author{Christiaan Verwijs}
\email{christiaan.verwijs@theliberators.com}
\affiliation{%
  \institution{The Liberators}
  \country{The Netherlands}
}

\author{Evelien Acun-Roos}
\email{eroos@xebia.com}
\affiliation{%
  \institution{Xebia}
  \country{The Netherlands}
}

\author{Daniel Russo}
\authornote{Corresponding author.}
\email{daniel.russo@cs.aau.dk}
\orcid{0000-0001-7253-101X}
\affiliation{%
  \institution{Department of Computer Science, Aalborg University}
  \streetaddress{A.C. Meyers Vaenge, 15, 2450}
  \city{Copenhagen}
  \country{Denmark}}

\renewcommand{\shortauthors}{Verwijs et al., 2025}

\begin{abstract}

As hybrid, distributed, and asynchronous work models become more prevalent, continuous learning in Agile Software Development (ASD) gains renewed importance. Communities of Practice (CoPs) are increasingly adopted to support social learning beyond formal education, often relying on virtual communication. Psychological safety, a prerequisite for effective learning, remains insufficiently understood in these settings. This mixed-methods study investigates psychological safety within Agile CoPs through survey data from 143 participants. Results indicate that psychological safety is significantly lower in online interactions compared to face-to-face settings. Moreover, low psychological safety reduces participants' intent to continue contributing and avoidance of interpersonal risk. No significant differences emerged based on gender, community seniority, or content creation activity. However, differences by role and age group suggest potential generational or role-related effects. Thematic analysis revealed exclusionary behavior, negative interaction patterns, and hostility as primary threats to psychological safety, often reinforced by tribalism and specific community dynamics. Suggested interventions include establishing explicit norms, structured facilitation, and active moderation. The findings were validated through member checking with 30 participants. This study provides a comparative perspective on interaction modalities and offers practical guidance for organizers seeking to cultivate inclusive, high-impact CoPs and similarly structured virtual or hybrid work environments.
\end{abstract}



\maketitle

\section{Introduction}
\label{sec:Introduction}

The ongoing transformation of work—toward hybrid, AI-supported, distributed, and asynchronous modes—challenges how knowledge is created, shared, and sustained. In Agile Software Development (ASD), a field marked by rapid change and evolving practices, the capacity for continuous learning is no longer optional—it is foundational~\cite{storey2014r}. Yet formal curricula often fail to keep pace. As a result, practitioners increasingly turn to Communities of Practice (CoPs) — defined by Wenger~\cite{wenger2009communities} as ``groups of people who share a concern or a passion for something they do and learn how to do it better as they interact regularly'' — to fill this gap. These communities, whether structured as meetups, chat groups, or forums, provide informal spaces where learning is peer-driven and iterative.

While the importance of CoPs has grown, particularly in distributed and online-first environments, the mechanisms that enable them to function effectively online are strained. One critical enabler — psychological safety — is at risk. When much of the interaction occurs through asynchronous messaging, video calls, or impersonal forums, traditional cues for trust and inclusion fade. This raises a pressing question: \textbf{How can psychological safety be maintained in virtual and hybrid CoPs, where shared norms and personal rapport are harder to establish?}

Psychological safety — defined as the belief that one can speak up, ask questions, and share ideas without fear of embarrassment or exclusion — is known to support learning, collaboration, and innovation~\cite{edmondson1999psychological,edmondson2023psychological}. However, the dynamics that shape it in loosely structured, voluntary, and often anonymous online settings remain largely unexplored. Participants in such communities are frequently unknown to one another, operate within flattened hierarchies, and may be subject to public scrutiny—all of which heighten perceived interpersonal risk. This study is motivated by three key gaps in existing knowledge:

\begin{enumerate}
\item A lack of empirical comparison between psychological safety in virtual and face-to-face CoPs.
\item Limited understanding of how psychological safety varies across demographic or professional categories such as age, gender, role, or seniority~\cite{edmondson2023psychological}.
\item Uncertainty about the feasibility and effectiveness of psychological safety interventions in decentralized, large-scale communities~\cite{o2020systematic}.
\end{enumerate}

To address these issues, we conducted a two-phase mixed-methods study. In Phase I, we surveyed members of Agile CoPs ($N = 143$), combining quantitative measures of psychological safety across modalities with qualitative insights from open-ended responses. In Phase II, we used member checking ($N = 30$) to validate our findings, elicit community interpretations, and co-create practical recommendations.

The following questions guide our research:

\begin{itemize}
\item RQ$_1$: \textit{How does psychological safety in Communities of Practice compare between interactions in different modalities (virtual or face-to-face)?}
\item RQ$_2$: \textit{How does psychological safety impact participants in Communities of Practice, and how can it be enhanced?}
\end{itemize}

Our findings show that psychological safety is significantly lower in online interactions, with notable variation across age groups and professional roles. Thematic analysis identified exclusionary behaviors, unconstructive interaction patterns, and hostility as central threats. Participants reported a reduction of contributions, avoidance of interpersonal risks, and emotional reactions as the most common responses. Tribalism and community dynamics further amplify these risks. In response, participants proposed countermeasures, including articulating shared norms, introducing structured facilitation, and implementing active moderation. \textbf{These results offer urgently needed guidance for practitioners aiming to sustain inclusive and effective learning environments in distributed contexts.}

The remainder of the paper is structured as follows. Section~\ref{sec:related} reviews prior work on Communities of Practice and psychological safety. Section~\ref{sec:researchdesign} describes our methodological approach. Sections~\ref{sec:phase1} and~\ref{sec:phase2} present findings from the survey study and the member checking process, respectively. We conclude by discussing implications and limitations (Section~\ref{sec:Discussion}) and outlining directions for future research (Section~\ref{sec:Conclusion}).

\section{Related work}
\label{sec:related}
The concept of Communities of Practice (CoPs) as a model for knowledge sharing and collaborative learning was introduced by Lave and Wenger~\cite{lave1991situated}. Departing from earlier cognitive theories of learning, they framed learning as a fundamentally social process, grounded in participation within a community of practitioners. Newcomers initially engage peripherally, gradually moving toward full participation as they develop expertise — a dynamic described as Legitimate Peripheral Participation (LPP)~\cite{wenger2009communities}. Wenger et al.~\cite{wenger2002cultivating} further formalized CoPs as entities characterized by a \textit{shared domain} of interest, a community with \textit{social bonds}, and a \textit{practice} encompassing tools, language, and routines.

\subsection{Prevalence of Communities of Practice}
CoPs are now widely established in domains such as healthcare~\cite{li2009use,andrew2008building}, software engineering~\cite{storey2014r}, education~\cite{hodgkinson2008developing}, and business~\cite{aljuwaiber2016communities}. While many are intra-organizational, or organizational Communities of Practice (oCoPs)~\cite{kimble2001communities}, they increasingly extend across institutional and geographic boundaries. Enabled by digital platforms like LinkedIn, Slack, and Zoom~\cite{preece2001sociability,storey2014r}, virtual CoPs mirror their in-person counterparts in domain focus, shared norms, and interactional dynamics~\cite{wenger2002cultivating}.

Agile software development (ASD) is a particularly fertile domain for CoPs. Meetup hosts over 300 face-to-face and virtual communities of practice focusing on Agile (approx. 200.000 members)~\cite{MeetupAgileScrum}, and over 960 that focus on software development (approx. 780.000 members)~\cite{meetup_software_engineering}. LinkedIn has over 6.200 discussion groups where members discuss challenges, questions, and experiences regarding Agile software development~\cite{LinkedInAgileGroups}. CoPs are also leveraged as part of organizational change initiatives in software firms~\cite{paasivaara2014communities,paasivaara2014deepening}.

\subsection{Antecedents for effective Communities of Practice}
The prevalence of Communities of Practice has led to substantial research on the factors that contribute to their effectiveness. Lave \& Wenger~\cite{lave1991situated} hypothesize in their Legitimate Peripheral Participation theory (LPP)~\cite{lave1991situated} that Communities of Practice are effective when learners see them as valuable repositories of knowledge (\textit{legitimacy}), when members feel integrated with them (\textit{peripherality}), and when Communities of Practice encourage active participation through a clear and shared purpose (\textit{participation}). 

Several empirical studies have identified antecedents of effective Communities of Practice. At the community level, trust, social identity, and reciprocity are repeatedly found to support knowledge sharing~\cite{kimble2001communities,wasko2000one,chiu2006understanding}. At the individual level, self-efficacy, community interest, and pro-social behavior are positively associated with engagement~\cite{ardichvili2003motivation,hsu2007knowledge}. Structural factors — such as member interdependence, managerial support, and density of social connections — also play a role, particularly in OCoPs~\cite{kirkman2013global,chen2007factors,smite2020spotify}. Preece~\cite{preece2001sociability} offers a framework for evaluating the usability and sociability of virtual communities, such as CoPs, and identifies a shared purpose, people, and good policies as key components.

\subsection{Psychological safety in Communities of Practice}
One factor with particular relevance to Communities of Practice is psychological safety. Edmondson~\cite[p. 9]{edmondson1999psychological} defines it as \emph{``a shared belief held by team members that the team is safe for interpersonal risk-taking''}, which draws on earlier work by the organizational theorist Schein~\cite{edmondson2014psychological}. It concerns explicitly the inter-personal risk of embarrassment or exclusion when voicing dissent, asking questions, or admitting ignorance~\cite{edmondson2004psychological}. While psychological safety is conceptually related to trust, the former is intended as a group-level phenomenon, whereas the latter is a characteristic of interpersonal relationships. Where trust is about giving another person the benefit of the doubt, psychological safety is about receiving the benefit of the doubt from  others in a group~\cite{edmondson1999psychological}. High psychological safety supports learning behaviors such as seeking feedback, offering ideas, and asking for help~\cite{edmondson1999psychological}, and has been linked to team effectiveness in software engineering and beyond~\cite{moe2010teamwork,strode2022teamwork,hennel2021investigating,verwijs2023theory}. Edmondson~\cite{edmondson2014psychological} emphasizes the need for constructive conflict to facilitate shared learning, though not in a way that ridicules, harms, or singles out individuals in a way that reduces their social standing in a group.

In CoPs, where participation is often voluntary and hierarchical boundaries are blurred, psychological safety seems especially critical. Zhang et al.~\cite{zhang2010exploring} found that psychological safety significantly predicted continued knowledge sharing in virtual CoPs. Similarly, Kirkman et al.~\cite{kirkman2013global} showed it buffered the negative effects of national diversity in global CoPs. However, psychological safety often relies on subtle nonverbal cues — tone, body language, facial expression — that are attenuated or absent in virtual settings~\cite{kostovich2020establishing,hayashida2024privacy,storey2014r}. This raises questions about whether psychological safety can be sustained online:

Hypothesis 1 (H1). \textit{Psychological safety is lower in virtual interactions than in face-to-face interactions.}\\

\subsection{Antecedents of psychological safety}
The vital role of psychological safety in learning, as seen in Communities of Practice, has led researchers to investigate its potential antecedents. Edmondson \& Bransby~\cite{edmondson2023psychological} note that demographic moderators, such as gender, age, and role, are rarely studied. Some studies have found small  effects~\cite{roh2022gender,cole2023impact,giordano2018psychological,ito2022concept}, but findings are inconsistent. For virtual CoPs, Zhang et al.~\cite{zhang2010exploring} found no significant effects of age or gender. The seniority of a member in their community, as well as whether they actively create content for it, may also shape perceptions of safety, as active or long-term contributors may be more exposed to scrutiny or feedback. As research to date has not shown substantial effects of demographic variables, we hypothesize:\\

Hypothesis 2 (H2). \textit{Psychological safety is not significantly different between genders (H2a), age groups (H2b), content creatorship (H2c), roles (H2d) and seniority (H2e).}\\

Beyond demographics, antecedents of psychological safety include leadership behaviors, interpersonal dynamics, and work design characteristics. Kahn~\cite{kahn1990psychological} identified interpersonal relationships and shared norms as foundational. Proactive personality traits and emotional stability have also been suggested~\cite{detert2007leadership,edmondson2006explaining}, though a meta-analysis showed mixed support for personality traits~\cite {frazier2017psychological}. Leadership style, cultural traits (e.g., uncertainty avoidance), and group autonomy have all been identified as influential moderators~\cite{frazier2017psychological,zboralski2009antecedents}.

\subsection{Interventions to improve psychological safety}
While psychological safety is widely discussed, evidence-based interventions remain limited. O'Donovan \& McAuliffe~\cite{o2020systematic} reviewed available interventions to improve psychological safety and found that most focused on education (case studies, workshops, simulation), with a few non-educational approaches (action research, forum play, and facilitation). The evidence for the efficacy of interventions was mixed, leading the authors to conclude that \emph{``there is a dearth of research on interventions that can be used to improve psychological safety or its related constructs.''}. Similar conclusions were drawn by Dusenberry \& Robinson~\cite{dusenberry2020building} in the education domain, and emphasized the need for contextualized approaches. Other studies highlight leadership messaging, codes of conduct, and trust-building exercises~\cite{hunt2021enhancing,roussin2008increasing}. Roleplay-based interventions have shown promise in some engineering teams~\cite{scarpinella2023can}, though results are inconsistent~\cite{campbell2024exploring}.

For virtual or large-scale CoPs, the feasibility of such interventions is unclear. Some scholars recommend establishing explicit behavioral norms and trust-building mechanisms~\cite{zhang2010exploring,lechner2022create}, but practical guidance and empirical validation are lacking. In short, while the importance of psychological safety is widely acknowledged, interventions tailored to distributed, voluntary communities remain scarce.

\section{Research Design}
\label{sec:researchdesign}
This study aims to understand how psychological safety impacts learning in Communities of Practice across different modalities. It consisted of two phases: a survey study and a validation phase. 

The first phase featured a survey study among members of Communities of Practice in Agile software engineering. The survey contained quantitative measures of psychological safety for online and face-to-face interactions. It also collected qualitative data about experiences with low safety, its effects on participants, contributing factors, and potential improvement strategies. The combination of quantitative and qualitative measures is described as Mixed-Method research~\cite{creswell2017designing}. Researchers can perform hypothesis testing on quantitative data~\cite{goodwin2016research} and use qualitative data to gain a deeper understanding of a phenomenon~\cite {creswell2016qualitative}. Additionally, it enhances the validity of inferences by triangulating results from different data sources~\cite{greene1989toward}. The design for Phase~I is described in detail in Section~\ref{sec:phase1}.

Because psychological safety is a highly subjective experience, we wanted to validate our findings with participants through \textit{member checking} in the second phase. Lincoln \& Guba~\cite{lincoln1986research} describe member checking as a process where researchers share their findings with some or all participants for feedback~\cite{mckim2023meaningful}, and use their expert perspective to evaluate credibility and correct mistakes~\cite{thomas2017feedback}. Given the community-oriented nature of the study, this also provided an opportunity to leverage the expertise of participants in generating insights from the results and research questions. Therefore, we follow the approach to member checking outlined by McKim~\cite{mckim2023meaningful}. The design for Phase~II is described in detail in Section~\ref{sec:phase2}.

\begin{figure}[H]
\centering
\includegraphics[height=0.65in]{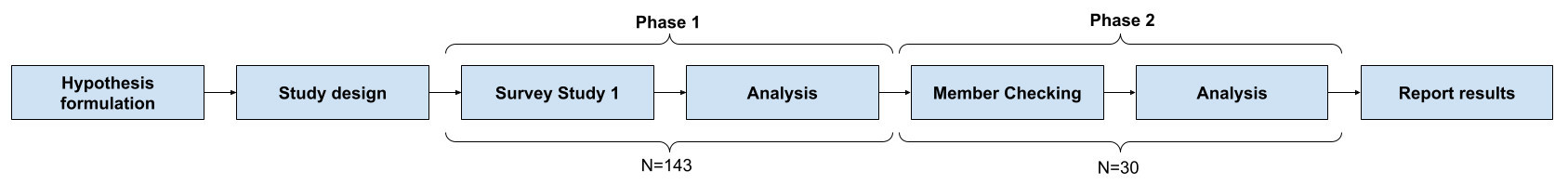}
\caption{The design of this two-phased study.}
\label{fig:steps}
\Description[The design of this two-phased study.]{This image shows the phases also described in this section.}
\end{figure}

The steps involved in our design are summarized in Figure \ref{fig:steps}. A replication package is available on Zenodo to encourage secondary studies. 

\section{Phase~I: Survey study}
\label{sec:phase1}

\subsection{Participants}
\label{sec:phase1participants}
We conducted our data collection process for Phase~II through an online survey deployed using IBM Qualtrics. Data was collected between December 2023 and March 2024. A purposive non-probabilistic sampling strategy~\cite{baltes2022sampling} was used. The strategy aimed to recruit individuals who were actively contributing to Communities of Practice in the Agile Software Development domain, as identified by the authors. This included several communities on LinkedIn, online and in-person meetup groups, professional conferences, relevant podcasts, and newsletters. Additionally, the authors reached out to several well-known community leaders to help promote the study and participate themselves.

The survey employed Likert scales to collect quantitative data and included essay questions to gather qualitative data. Appendix~\ref{sec:appendix} includes the complete survey (Table~\ref{tab:appendix-phase1-survey}). To reduce self-selection bias, especially among individuals with low psychological safety, we avoided collecting personally identifiable information (e.g., name, email address) and excluded subgroup analyses with fewer than three respondents~\cite{meade2012identifying}. We explained these protections in the survey introduction and all invitations.

We received 159 responses. To minimize response bias, we removed 16 responses that were completed within 180 seconds or were careless~\cite{meade2012identifying}. 143 participants were retained. The distribution in our sample (by gender, age group, content creatorship, role, and seniority) is shown in Table~\ref{tab:phase1samplecomposition}.

\begin{table}[H]
\centering
\tiny
\caption{Composition of the sample for Phase~I}
\label{tab:phase1samplecomposition}
\begin{tabular}{m{3.5cm}p{4cm}p{1.6cm}}
\toprule
\textbf{Variable} & \textbf{Category} & \textbf{N (\%)} \\ \midrule
Respondents & & 143 (100\%) \\
Gender & Women & 43 (30.1\%) \\
 & Men & 90 (62.9\%) \\
 & Other / Prefer not to say & 10 (7.0\%) \\
Age group & 26-35 years & 14 (9.8\%) \\
 & 36-45 years & 44 (30.8\%) \\
 & 46-55 years & 54 (37.8\%) \\
 & 56-65 years & 21 (14.7\%) \\
 & 66+ years & 5 (3.5\%) \\
 & Prefer not to say & 5 (3.5\%) \\
Role & Scrum Master & 35 (24.5\%) \\
 & Agile Coach & 49 (34.3\%) \\
 & Trainer or facilitator & 17 (11.9\%) \\
 & Consultant & 21 14.7\%) \\
 & Other & 18 (12.6\%) \\
 & Prefer not to say & 3 (2.1\%) \\
Seniority in community & Less than 2 years & 3 (2.1\%) \\
 & Between 2 and 5 years & 14 (9.8\%) \\
 & Between 5 and 10 years & 56 (39.2\%) \\
 & More than 10 years & 70 (49.0\%) \\
Content creator & No & 56 (39.2\%) \\ 
 & Yes & 87 (60.8\%) \\
 \bottomrule
\end{tabular}
\end{table}

\subsection{Measurements}
\label{sec:results-phase1-measurements}
\textbf{Psychological safety:} We asked participants to evaluate the level of psychological safety they experienced in their Community of Practice with a psychometric Likert scale (1–5). This scale adapted four items from Edmondson’s original measure~\cite{edmondson1999psychological}, including ``Members of the Agile community are open to different perspectives and opinions'' and ``Members of the Agile community foster an environment where it is easy to ask for help.'' We added one item to assess the perceived importance of psychological safety (``It is important to me that interactions in the Agile community are psychologically safe') and another to test face validity (``I experience psychological safety in the Agile community to be high'').

Participants completed this scale once for the face-to-face modality and once for the virtual modality. For virtual modalities, participants were asked to keep the online medium they used the most in mind (e.g., LinkedIn, Slack, Discord, forums). This recognizes that the level of psychological safety may depend on the medium used, with discussion threads on LinkedIn or fora being more public, whereas Slack and Discord are often members-only.

We randomized the order of items to reduce order bias. The reliability of the original scale was high for both modalities (face-to-face: $\alpha=.86$; online: $\alpha=.85$). However, the item included to assess the importance of psychological safety (``It is important to me that interactions in the Agile community are psychologically safe') showed a low item-total correlation (ITC), indicating separate factors. The item was removed. The item included to test face validity (``I experience psychological safety in the Agile community to be high'') showed excellent ITC for both modalities ($r = .82$ and $r = .84$, respectively), indicating good face validity, and was retained in both scales.
The resulting measurement reliability was high for face-to-face interactions ($\alpha=.89$) and online interactions ($\alpha=.89$). A Confirmatory Factor Analysis (CFA) showed that the remaining items loaded on their expected factor. Together, the two factors explained 60.6\% of total variance. The two factors were moderately correlated ($r = .45$, $N=130$).

\textbf{Qualitative questions:} Four essay questions were included to explore the qualitative experience of low psychological safety, its effects on participants, and strategies to improve it. 
\begin{itemize}
    \item ``What are examples of behavior you have personally encountered during interactions in the Agile community that lowered psychological safety for you?''
    \item ``How does a lack of psychological safety in interactions in the Agile community affect you or your behavior?''
    \item ``In your experience, what behavior or factors decrease psychological safety in the Agile community?''
    \item ``What measures or strategies could improve psychological safety and encourage more respectful and constructive discussions within the Agile community?''
\end{itemize}

We did not distinguish between modalities for these questions to avoid overloading participants. We analyzed responses using the qualitative coding approach described in Section~\ref{sec:results-phase1-analysis-qualitative}.\\

\textbf{Gender:} The gender of participants was measured with a single categorical item ``What is your gender?''. The available options were ``man'', ``woman'', ``non-binary / third gender'' and ``prefer not to say''.

\textbf{Age group:} The age group of participants was measured with a single categorical item (``What is your age group?'') that invited participants to self-categorize into 10-year buckets: ``18-25 years'', ``26-35 years'', ``36-45 years'', ``46-55 years'', ``56-65 years'', ``65+ years'' and ``prefer not to say''.

\textbf{Role:} For role, we asked participants to self-categorize into the most applicable role (``What best describes your role?''). Because the study was run in Communities of Practice related to Agile and software development, the available options fitted those communities: ``developer, analyst, tester, designer'', ``scrum master'', ``agile coach'', ``trainer or facilitator'', ``consultant'', ``other'' and ``Prefer not to say''.

\textbf{Seniority:} Seniority in the community was measured through a single categorical item ``How many years have you been active in the Agile community?''. The available options were ``Less then a year'', ``Between 1 and 2 years'', ``Between 2 and 5 years'', ``Between 5 and 10 years'' and ``More than 10 years''.

\textbf{Role:} For role, we asked participants to self-categorize into the most applicable role (``What best describes your role?''). Because the study was run in Communities of Practice related to Agile and software development, the available options fitted those communities: ``developer, analyst, tester, designer'', ``scrum master'', ``agile coach'', ``trainer or facilitator'', ``consultant'', ``other'' and ``Prefer not to say''.

\textbf{Content creatorship:} To identify content creators in the sample, a single categorical question was included to evaluate it: ``Do you create content for the Agile community (videos, blogs, products, etc?)''. Available options were ``Yes''; and ``No''.

Seniority in the community was measured through a single categorical item ``How many years have you been active in the Agile community?''. The available options were ``Less then a year'', ``Between 1 and 2 years'', ``Between 2 and 5 years'', ``Between 5 and 10 years'' and ``More than 10 years''.

\textbf{Interactions with Communities of Practice:} Several questions were included to gain insight into where and how participants interacted with Communities of Practice. The first was a set of nine categorical questions that asked participants to evaluate how often they shared something with Agile Communities of Practice on the following online platforms: ``LinkedIn'', ``Twitter / X'', ``Medium'', ``Mastodon'', ``Slack / Discord'', ``Facebook'', ``forums'' and ``YouTube''. The available platforms were chosen by the first two authors of the paper, who are themselves active in the Agile community. The available options were ``at least once per day'', ``at least once per week'', ``at least once per month'', ``less than once a month'' and ``never''. Similarly, participants were asked to answer how often they interacted with Agile Communities of Practice in face-to-face settings. The available settings were ``conferences'', ``online meetups'', ``meetups'', ``workshop or training'' and ``other''. ``online meetups'' were included here because we felt them to be more face-to-face than the more anonymous interactions in online environments. For the frequency of interaction, the available categories were ``very often'', ``often'', ``rarely'' and ``never''. We decided not to use interval-specific options like those for online interactions because it is unlikely that participants are logistically able to attend conferences, training, or meetups every day, week, month, and so on.

\subsection{Analysis of the Qualitative Data}
\label{sec:results-phase1-analysis-qualitative}
Qualitative data was analyzed following a naturalistic inquiry paradigm through constant comparison~\cite{glaser2017discovery}. Thematic analysis was used to iteratively identify, refine, and report themes in thick qualitative data~\cite{terry2017thematic,braun2006using} and induce a conceptual model for the experience of psychological safety. To reinforce scientific rigor and increase replicability, we used the structured approach recommended by Gioia et. al.~\cite{gioia2013seeking} as an analytical framework. This framework is increasingly common in qualitative research due to its methodological clarity, traceability, and transparency~\cite{guba1981criteria}. 
The Gioia framework aims to develop \textit{data structures} from qualitative data through three stages. In the first stage, open coding was employed to analyze the responses to each essay question, identifying one or more first-order concepts that closely align with the language used by participants. In the second stage, axial coding was used to aggregate first-order concepts into related second-order themes, which were grouped into top-level aggregate dimensions in the third stage. The first and second stages are performed with as little theoretical knowledge as possible to prevent confirmation bias~\cite{gioia2013seeking}. A total of 11.415 words were analyzed and coded.
To strengthen validity, multiple coders were involved in the coding process. The study's first author coded the cases and developed an initial codebook in the software package QualCoder~\cite{curtain2025qualcoder}. The study's second author independently coded all cases with this codebook. The corresponding author supervised and assessed the overall correctness of the process. Emerging codes, themes, and dimensions were memoed as the coders iterated until no further insights could be induced~\cite{saunders2018saturation}. 

Cohen's Kappa was calculated for each first-order concept to assess inter-rater reliability and identify areas of improvement ($\kappa < .60$). For the four main themes, coding convergence was achieved with high agreement:
\begin{itemize}
    \item \textit{Experiences of low psychological safety:} 51 first-order concepts, 4 iterations, $\kappa = .92$
    \item \textit{Effects of low psychological safety:} 67 first-order concepts, 2 iterations, $\kappa = .94$
    \item \textit{Factors contributing to low psychological safety:} 67 first-order concepts, 2 iterations, $\kappa = .93$
    \item \textit{Improvement strategies:} 54 first-order concepts, 2 iterations, $\kappa = .95$
\end{itemize}

Inter-rater reliability was also assessed for second-order themes and thematic groupings, which ranged from .77 to 1.0, indicating high consistency between raters. This iterative process resulted in a thematic map that outlines the experience of psychological safety in Communities of Practice.

\subsection{Analysis of the Quantitative Data}
\label{sec:results-phase1-analysis-quantitative}
The quantitative data was analyzed using IBM SPSS. Outlier detection was performed with stem-and-leaf plots. Five outliers were identified for the reported psychological safety in face-to-face interactions, but none were found for online interactions. Since the outliers appeared to be legitimate cases with low safety rather than data errors, we retained them. However, we reran statistical tests without outliers and report differences where relevant. 

We checked for missing data and found between two and three cases missing per continuous variable. Little’s MCAR test confirmed the data were missing completely at random ($\chi^2 = 27.091$, $df = 28$, $p = 0.53$)~\cite{hair2019multivariate}. We used listwise deletion where relevant.

To compare psychological safety across modalities, we used a paired samples t-test. We confirmed the assumptions of normality through kurtosis ($<3$), skewness ($<2$), and Q-Q plots~\cite{hair2019multivariate}. However, a Kolmogorov-Smirnov test revealed a significant deviation from normality, which could introduce bias in our estimates. As recommended in statistical literature, bootstrapping was used to normalize the distributions and reduce bias in the resulting estimates~\cite{efron1992bootstrap}.

We used ANOVA to test for group differences by gender, age, role, seniority, and content creation, both for quantitative scale scores and for frequencies of qualitative codes. Because Levene’s test indicated unequal variances, we used Welch’s ANOVA with Games-Howell post hoc comparisons, which maintain robustness without assuming homogeneity~\cite{brown1974robust}. A 90\% confidence interval was calculated for the effect sizes based on the approach outlined by Smithson~\cite{smithson2003confidence}. For paired 

We performed a post hoc power analysis using G*Power~\cite{faul2009statistical}, version 3.1.9. We determined that our sample for Phase~II allows us to correctly capture small effects ($\textit{d}=.020$) with a statistical power of ~84\% ($1-\beta= .840$) for the sample of 143 participants, and a power of 99.8\% ($1-\beta= 1.000$) for moderate effects ($\textit{d}=.050$). Thus, we are confident that we can detect moderate effects and most small effects.

We report effect sizes throughout this study in addition to their significance. The use of effect sizes requires researchers to interpret the degree to which the results diverge from expectations as a measure of this meaningfulness~\cite{vacha2004estimate,kelley2012effect}. For paired t-tests, we calculate Hedges' g as described by Smithson~\cite{smithson2003confidence}. It calculates the distance between two distributions of within-subject data. For Analysis of Variance, we calculate the eta-squared ($\eta^2$) as described in Ellis~\cite{ellis2010essential}. This statistic represents the amount of variance in the dependent variable that is explained by the independent variables. For both types of effect sizes, a 90\% confidence interval was calculated based on Smithson~\cite{smithson2003confidence}. Effect sizes below .01 indicate no effect. Effect sizes of 0.02, 0.06, and 0.14 are considered small, moderate, and large in magnitude, respectively. 

\subsection{Results}

\subsubsection{Where people interact}
This study examines psychological safety in Communities of Practice for different modalities. Table~\ref{tab:phase1-wherewemeet-online} shows where and how often participants shared content with their Community of Practice online. Although interactions happen across these platforms, LinkedIn, Slack / Discord, and forums are relatively the most popular, with respectively 33.8\%, 23.3\% and 7.9\% contributing at least weekly.

\begin{table}[H]
\centering
\caption{Frequency by which participants share with their Community of Practice in online settings (N=143).}
\label{tab:phase1-wherewemeet-online}
\begin{tabular}{m{3cm}p{1.5cm}p{1.5cm}p{1.5cm}p{1.5cm}p{1.5cm}}
\toprule
\textbf{Platform} & \textbf{Never} & \textbf{Less than monthly} & \textbf{Monthly} & \textbf{Weekly} &\ \textbf{Daily} \\ \midrule
LinkedIn & 10.0\% & 32.3\% & 23.8\% & 23.1\% & 10.8\% \\
Slack / Discord & 47.3\% & 17.8\% & 11.6\% & 17.8\% & 5.4\% \\
Forums & 57.1\% & 26.4\% & 8.6\% & 7.9\% & 0\% \\
Twitter / X & 69.8\% & 18.6\% & 3.9\% & 7.0\% & .8\% \\
Facebook & 73.0\% & 15.6\% & 4.3\% & 5.7\% & 1.4\% \\
YouTube & 68.8\% & 19.9\% & 7.8\% & 2.8\% & .7\% \\
Medium & 74.4\% & 18.6\% & 3.9\% & 2.3\% & .8\% \\
Reddit & 85.2\% & 10.9\% & .8\% & 2.3\% & .8\% \\
Mastodon & 93.0\% & 2.3\% & 1.6\% & 1.6\% & 1.6\% \\
 \bottomrule
\end{tabular}
\end{table}

The frequency of interactions in face-to-face settings is reported in Table~\ref{tab:phase1-wherewemeet-facetoface}. Of the settings provided, workshops and training, online meetups, and conferences are the most popular, with 55.6\%, 57.0\% and 42.7\% reporting at least ``often''.

\begin{table}[H]
\centering
\caption{Frequency by which participants interact with their Community of Practice in face-to-face settings (N=143).}
\label{tab:phase1-wherewemeet-facetoface}
\begin{tabular}{m{3cm}p{1.5cm}p{1.5cm}p{1.5cm}p{1.5cm}}
\toprule
\textbf{Platform} & \textbf{Never} & \textbf{Rarely} & \textbf{Often} & \textbf{Very often} \\
\midrule
Workshop or training & 4.2\% & 40.1\% & 38.0\% & 17.6\% \\
Online meetups & 10.6\% & 32.4\% & 44.4\% & 12.7\% \\
Conferences & 12.0\% & 45.1\% & 32.4\% & 10.6\% \\
Other & 21.8\% & 38.7\% & 29.0\% & 10.5\% \\
In-person meetups & 7.8\% & 53.2\% & 28.4\% & 10.6\% \\
 \bottomrule
\end{tabular}
\end{table}

\subsubsection{Group differences in quantitative measures of psychological safety}
Table ~\ref{tab:results-phase1-groupdifferences} shows the within-subject and between-subject differences for face-to-face and online modalities and between demographic variables.

\begin{table}[H]
\centering
\tiny
\caption{Within-subject and between-subject differences in reported psychological safety between genders, age groups, content creatorship, role, and seniority. Statistical group differences ($p < .05$) are marked with *. Effect sizes with 90\% confidence intervals are included, with Hedges' g for within-subject differences and eta squared ($\eta^2$) for between-subject differences.}
\label{tab:results-phase1-groupdifferences}
\begin{tabular}{m{2.8cm}p{0.5cm}p{0.5cm}p{0.5cm}p{0.5cm}p{1.5cm}p{0.5cm}p{0.5cm}p{0.5cm}p{0.5cm}p{1.5cm}}
\toprule
\textbf{Variable} & \multicolumn{5}{c}{\textbf{Face-to-face (N=141)}} & \multicolumn{5}{c}{\textbf{Online (N=143)}} \\
\midrule
 & \textbf{Mean} & \textbf{s} & \multicolumn{1}{l}{} & \textbf{} & \textbf{} & \textbf{Mean} & \textbf{s} & \textbf{} & \textbf{p} & \textbf{Hdgs' g / 90\% CI} \\
\midrule
All (N=141) & 3.879 & .880 & \multicolumn{1}{l}{} & \multicolumn{1}{l}{} & \multicolumn{1}{l}{} & 2.910 & 1.027 &  & .000* & 1.008 (.802, 1.222) \\
\midrule
\textbf{By gender} & \textbf{Mean} & \textbf{s} & \textbf{F} & \textbf{p} & \textbf{$\eta^2$ / 90\% CI} & \textbf{Mean} & \textbf{S} & \textbf{F} & \textbf{p} & \textbf{$\eta^2$ / 90\% CI} \\
\midrule
Women (N=43) & 3.735 & .816 & 2.717 & .070 & .039 (.000, .094) & 2.870 & 1.133 & .921 & .401 & .013 (.000, .050) \\
Men (N=90) & 4.000 & .872 &  &  &  & 3.001 & 1.003 &  &  &  \\
Other / Prefer not to say (N=10) & 3.440 & 1.070 &  &  &  & 2.520 & .755 &  &  &  \\
\midrule
\textbf{By age group} & \textbf{Mean} & \textbf{s} & \textbf{F} & \textbf{p} & \textbf{$\eta^2$ / 90\% CI} & \textbf{Mean} & \textbf{S} & \textbf{F} & \textbf{p} & \textbf{η2 / 90\% CI} \\
\midrule
26-35 years (N=14) & 3.857 & .721 & 1.946 & .091 & .072 (.000, .113) & 3.314 & 1.106 & 3.714 & .004* & .138 (.025, .182) \\
36-45 years (N=44) & 3.845 & .925 &  &  &  & 2.659 & 1.006 &  &  &  \\
46-55 years (N=54) & 3.992 & .894 &  &  &  & 2.871 & .990 &  &  &  \\
56-65 years (N=21) & 3.890 & .666 &  &  &  & 3.429 & .823 &  &  &  \\
66+ years (N=5) & 4.120 & 1.026 &  &  &  & 3.720 & 1.205 &  &  &  \\
Prefer not to say (N=5) & 2.760 & .932 &  &  &  & 1.920 & .576 &  &  &  \\
\midrule
\textbf{By content creatorship} & \textbf{Mean} & \textbf{s} & \textbf{F} & \textbf{p} & \textbf{$\eta^2$ / 90\% CI} & \textbf{Mean} & \textbf{s} & \textbf{F} & \textbf{p} & \textbf{$\eta^2$ / 90\% CI} \\
\midrule
No (N=56) & 3.926 & .870 & .243 & .623 & .002 (.000, .030) & 3.000 & .944 & .174 & .677 & .001 (.000, .027) \\
Yes (N=87) & 3.851 & .890 &  &  &  & 2.881 & 1.086 &  &  &  \\
\midrule
\textbf{By role} & \textbf{Mean} & \textbf{s} & \textbf{F} & \textbf{p} & \textbf{$\eta^2$ / 90\% CI} & \textbf{Mean} & \textbf{s} & \textbf{F} & \textbf{p} & \textbf{$\eta^2$ / 90\% CI} \\
\midrule
Scrum Master (N=35) & 4.035 & .906 & 2.541 & .031* & .094 (.000, .138) & 3.234 & .839 & 2.120 & .067 & .079 (.000, .121) \\
Agile Coach (N=49) & 3.875 & .913 &  &  &  & 2.883 & 1.100 &  &  &  \\
Trainer or facilitator (N=17) & 3.529 & .900 &  &  &  & 2.659 & .908 &  &  &  \\
Consultant (N=21) & 4.095 & .571 &  &  &  & 2.886 & .956 &  &  &  \\
Other (N=18) & 3.900 & .762 &  &  &  & 3.000 & 1.240 &  &  &  \\
Prefer not to say (N=3) & 2.533 & 1.332 &  &  &  & 1.467 & .231 &  &  &  \\
\midrule
\textbf{By seniority} & \textbf{Mean} & \textbf{s} & \textbf{F} & \textbf{p} & \textbf{$\eta^2$ / 90\% CI} & \textbf{Mean} & \textbf{s} & \textbf{F} & \textbf{p} & \textbf{$\eta^2$ / 90\% CI} \\
\midrule
Less than 2 years (N=3) & 4.533 & .643 & .700 & .554 & .015 (.000, .044) & 3.333 & .416 & .413 & .744 & .009 (.000, .029) \\
Between 2 and 5 years (N=14) & 3.957 & .816 &  &  &  & 3.114 & 1.089 &  &  &  \\
Between 5 and 10 years (N=56) & 3.900 & .904 &  &  &  & 2.947 & 1.131 &  &  &  \\
More than 10 years (N=70) & 3.820 & .885 &  &  &  & 2.857 & .959 &  &  & \\
 \bottomrule
\end{tabular}
\end{table}

First, psychological safety is higher ($M=3.879, N=141$) in face-to-face interactions than in online interactions ($M=2.910, N=143$). The difference was statistically significant ($p < .001$) and supports Hypothesis 1. The overall effect size is qualified as \textit{large}, $Hedges' g = 1.008, 90\% CI [.802, .1.222]$

The results for group differences showed mixed results. No significant gender difference exists for face-to-face interactions ($p < .05$), with a \textit{small} effect size ($\eta^2 = .039, 90\% CI [.000, .094]$). There are also no gender differences for online interactions ($p < .05$), with an effect size classified as \textit{small}, $\eta^2 = .013, 90\% CI [.000, .050]$. Thus, the results are consistent with Hypothesis H2a.

For age, psychological safety was not statistically different between age groups for face-to-face interactions ($p < .05$). However, a significant difference was found between age groups for online interactions ($p < .01$). Moreover, the effect size for both contexts was respectively \textit{moderate} ($\eta^2 = .072, 90\% CI [.000, .113]$) and \textit{large} ($\eta^2 = .138, 90\% CI [.025, .182]$). A small group of participants who preferred not to disclose their age reported very low safety for face-to-face ($M=2.760, N=5$) and online ($M=1.920, N=5$) interactions. For online interactions, psychological safety starts moderately high ($M=3.311, N=14$) for the 26-35 age group, then drops to the lowest level for the age group 36-45 ($M=2.659, N=44$) and incrementally increases as participants grow older, with the highest psychological safety reported by the age group of 66+ ($M=3.720, N=5$). Thus, our findings do not support hypothesis H2b.

In support of hypothesis H2c, no significant difference was found for content creatorship. The effect sizes are also classified as \textit{none}, respectively $\eta^2 = .002, 90\% CI [.000, .130]$ and $\eta^2 = .001, 90\% CI [.000, .027]$.

For roles, a significant difference was found for face-to-face interactions ($p < .031$), but not for online interactions ($p < .067$). The size of both effects was classified as \textit{moderate}, respectively $\eta^2 = .094, 90\% CI [.000, .138]$ and $\eta^2 = .079, 90\% CI [.000, .121]$. For face-to-face interactions, the highest safety is reported by ``Scrum Masters'' ($3.23, N=35$) and the lowest by trainers and facilitators ($2.66, N=21$). These findings do not support hypothesis H2d.

Finally, for seniority, no significant difference was found in reported psychological safety between different levels of seniority. The effect sizes are respectively classified as \textit{none} ($\eta^2 = .015, 90\% CI [.000, .044]$) and \textit{none} ($\eta^2 = .009, 90\% CI [.000, .029]$).

We now turn to the results from our qualitative analyses.

\subsubsection{Which behaviors decrease psychological safety in Communities of Practice?}
\label{sec:gioia-experience}
To better understand the qualitative experience of psychological safety, participants were asked to describe the behaviors that decreased psychological safety in their interactions. As described in Section~\ref{sec:results-phase1-analysis-qualitative}, responses were analyzed and coded into a structure of first-order concepts, second-order themes, and third-order aggregate dimensions with the Gioia methodology~\cite{gioia2012organizational}. The Gioia schema, along with the percentage of cases that reported each code, is shown in Figure~\ref{fig:gioia-experience} in Appendix~\ref{sec:appendix}.

The behaviors that decreased psychological safety in Communities of Practice manifested as four aggregate dimensions. The most reported dimension was ``Exclusionary behaviors'' (63.1\%), and consisted of the themes ``dogmatism'' (43.8\%) and ``dismissiveness'' (35.4\%). The second most-reported dimension was ``negative interaction patterns'' (33.8\%). It consisted of the themes ``unconstructive criticism'' (18.5\%), ``posturing'' (9.2\%), ``impoliteness'' (7.7\%), ``discrimination and sexism'' (5.4\%), and ``insensitivity'' (4.6\%). Next, 33.1\% of participants reported a code in the dimension ``hostile behavior'' (33.1\%), representing the themes ``personal attacks'' (20.8\%) and forms of ``aggression'' (14.6\%). The fourth dimension was ``unsafe environment'' (8.5\%), with 5.4\% of participants reporting behaviors contributing to unsafety like gossiping and a bad discussion culture, followed by ``negative use of humor'' (1.5\%) and ``unethical behavior'' (1.5\%). Finally, 9.2\% of participants reported no experience with behaviors that lowered psychological safety.

In summary, psychological safety is decreased primarily through exclusionary behaviors, where other members of a community dismiss, disregard, or disqualify contributions or the people who contribute them.

\subsubsection{The effect of low psychological safety on participants in Communities of Practice}\
\label{sec:gioia-effects}
Participants also described how low psychological safety impacted them in Communities of Practice. The Gioia schema, along with the percentage of cases that reported each code, is shown in Figure~\ref{fig:gioia-effects} in Appendix~\ref{sec:appendix}.

Six aggregate dimensions emerged from the qualitative coding. The most reported dimension was ``reduce contributions'' (58.5\%), consisting of the themes ``avoid or stop debates'' (23.8\%), ``share less'' (21.5\%), ``distance from community'' (12.3\%), ``decrease engagement'' (9.2\%) and ``leave parts of community'' (3.8\%). The second most-reported dimension was ``avoid interpersonal risks'' (20.0\%), representing the themes ``self-censorship'' (11.5\%), ``avoid situations'' (7.7\%) and ``stop asking questions'' (3.1\%). Next, 16.9\% of cases reported codes for the aggregate dimension ``emotional response'' (16.9\%), consisting of the themes ``Self-directed emotions'' (9.2\%) such as loneliness and anxiety, ``other-directed emotions'' (6.9\%) such as disappointment and annoyance, and ``loss of confidence'' (3.1\%). 10.8\% of respondents reported effects that reflected a desire to improve the situation, aggregated into the dimension ``improve situation'', with the underlying themes ``improve safety for others'' (5.4\%) and ``focus on the positive'' (6.9\%). Finally, 8.5\% of cases reported codes aggregated into the dimension  ``personal reflections'', consisting of the themes ``re-evaluate expectations''(3.1\%), ``recognize larger implications''(3.1\%) and  ``''(2.3\%). Finally, 11.5\% of participants reported ``no effect'' (11.5\%).

In summary, participants primarily reduce their contributions or interactions to Communities of Practice when faced with low psychological safety. Additionally, low psychological safety also creates a negative emotional response in many participants.

\subsubsection{Group differences in qualitative responses}\
We now turn to the differences in the qualitative experiences with psychological safety as reported by participants. Table~\ref{tab:results-phase1-groupdifferencesqualitative} shows the percentage of participants that reported each aggregate dimension identified in Section~\ref{sec:gioia-experience} and Section~\ref{sec:gioia-experience} for, respectively, behaviors that decreased their experience of psychological safety and the effect of low psychological safety on them. Most group differences are not significant. However, men reported a significantly lower percentage of negative interactions ($24.7\%$) than women ($47.5\%$) and other genders ($55.6\%$). More significant group differences were found between age groups, with the 46-55 cohort reporting the most exclusionary behaviors ($80\%$), the 65+ cohort reporting the most negative interactions ($75\%$), and the 26-35 cohort reporting the most hostile behaviors ($58.3\%$). 
A caveat with these results is that subgroups can be small, ranging from 51 to 4 members. Differences must be interpreted cautiously, particularly with smaller subgroups (see Table~\ref{tab:phase1samplecomposition}). However, we note that the findings are mostly consistent with our quantitative measure of psychological safety reported in Tables~\ref{tab:results-phase1-groupdifferencesqualitative-gender} through ~\ref{tab:results-phase1-groupdifferencesqualitative-role}, where few significant group differences were found, except for age and role in online settings. Thus, a reasonable conclusion from these results is that the overall experience of psychological safety among participants is relatively similar across subgroups.

\begin{table}[H]
\tiny
\caption{Gender differences in the percentage of participants that reported aggregate dimensions in behaviors that reduce psychological safety in Communities of Practice, and the effect on participants (N=143). Groups with fewer than 3 participants are suppressed to protect anonymity. Statistical group differences ($p < .05$) are marked with *.}
\label{tab:results-phase1-groupdifferencesqualitative-gender}
\begin{tabular}{lccccc}
\toprule
\textbf{Dimension} & \textbf{All} & \textbf{Women} & \textbf{Men} & \textbf{Other} & \textbf{p} \\
\midrule
\multicolumn{6}{l}{\textbf{Behaviors contributing to low psychological safety}} \\
\midrule
Negative interactions & 33.8\% & 47.5\% & 24.7\% & 55.6\% & .016 \\
Hostile behaviors & 33.1\% & 32.5\% & 33.3\% & 33.3\% & .996 \\
Exclusionary behaviors & 63.1\% & 65.0\% & 63.0\% & 55.6\% & .817 \\
Unsafe environment & 8.5\% & 7.5\% & 8.6\% &  & .937 \\
No experience & 9.2\% & 10.0\% & 8.6\% &  & .952 \\
\midrule
\multicolumn{6}{l}{\textbf{Effects of low psychological safety}} \\
\midrule
Emotional response & 16.9\% & 20.0\% & 16.0\% &  & .772 \\
Avoid interpersonal risks & 20.0\% & 25.0\% & 18.5\% &  & .560 \\
Personal reflections & 8.5\% & 10.0\% & 6.2\% &  & .242 \\
Reduce contributions & 58.5\% & 60.0\% & 58.0\% & 55.6\% & .963 \\
Improve situation & 10.8\% & 17.5\% & 6.2\% &  & .087 \\
No effect & 11.5\% & 10.0\% & 12.3\% &  & .931 \\
\bottomrule
\end{tabular}
\end{table}

\begin{table}[H]
\tiny
\caption{Age group differences in the percentage of participants that reported aggregate dimensions in behaviors that reduce psychological safety in Communities of Practice, and the effect on participants (N=143). Groups with fewer than 3 participants are suppressed to protect anonymity. Statistical group differences ($p < .05$) are marked with *.}
\label{tab:results-phase1-groupdifferencesqualitative-age}
\begin{tabular}{lccccccc}
\toprule
\textbf{Dimension} & \textbf{All} & \textbf{26-35} & \textbf{36-45} & \textbf{46-55} & \textbf{56-65} & \textbf{65+} & \textbf{p} \\
\midrule
\multicolumn{8}{l}{\textbf{Behaviors contributing to low psychological safety}} \\
\midrule
Negative interactions & 33.8\% & 41.7\% & 39.0\% & 18.0\% & 36.8\% & 75.0\% & .003 \\
Hostile behaviors & 33.1\% & 58.3\% & 24.4\% & 44.0\% &  &  & .011 \\
Exclusionary behaviors & 63.1\% & 58.3\% & 51.2\% & 80.0\% & 57.9\% &  & .006 \\
Unsafe environment & 8.5\% & 25.0\% & 9.8\% &  &  &  & .113 \\
No experience & 9.2\% &  & 9.8\% &  & 21.1\% &  & .010 \\
\midrule
\multicolumn{8}{l}{\textbf{Effects of low psychological safety}} \\
\midrule
Emotional response & 16.9\% &  & 19.5\% & 18.0\% &  &  & .884 \\
Avoid interpersonal risks & 20.0\% & 25.0\% & 22.0\% & 18.0\% &  &  & .560 \\
Personal reflections & 8.5\% &  & 12.2\% & 6.0\% &  &  & .686 \\
Reduce contributions & 58.5\% & 50.0\% & 63.4\% & 58.0\% & 68.4\% &  & .397 \\
Improve situation & 10.8\% &  & 7.3\% & 12.0\% &  &  & .808 \\
No effect & 11.5\% &  & 17.1\% & 6.0\% &  &  & .402 \\
\bottomrule
\end{tabular}
\end{table}

\begin{table}[H]
\tiny
\caption{Group differences in content creatorship in the percentage of participants that reported aggregate dimensions in behaviors that reduce psychological safety in Communities of Practice, and the effect on participants (N=143). Groups with fewer than 3 participants are suppressed to protect anonymity. Statistical group differences ($p < .05$) are marked with *.}
\label{tab:results-phase1-groupdifferencesqualitative-contentcreatorship}
\begin{tabular}{lcccc}
\toprule
\textbf{Dimension} & \textbf{All} & \textbf{Yes} & \textbf{No} & \textbf{p} \\
\midrule
\multicolumn{5}{l}{\textbf{Behaviors contributing to low psychological safety}} \\
\midrule
Negative interactions & 33.8\% & 34.6\% & 32.7\% & .825 \\
Hostile behaviors & 33.1\% & 38.3\% & 24.5\% & .107 \\
Exclusionary behaviors & 63.1\% & 67.9\% & 55.1\% & .145 \\
Unsafe environment & 8.5\% & 9.9\% & 6.1\% & .460 \\
No experience & 9.2\% & 11.1\% & 6.1\% & .345 \\
\midrule
\multicolumn{5}{l}{\textbf{Effects of low psychological safety}} \\
\midrule
Emotional response & 16.9\% & 14.8\% & 20.4\% & .414 \\
Avoid interpersonal risks & 20.0\% & 19.8\% & 20.4\% & .929 \\
Personal reflections & 8.5\% & 7.4\% & 10.2\% & .582 \\
Reduce contributions & 58.5\% & 54.3\% & 65.3\% & .221 \\
Improve situation & 10.8\% & 9.9\% & 12.2\% & .676 \\
No effect & 11.5\% & 12.3\% & 10.2\% & .714 \\
\bottomrule
\end{tabular}
\end{table}

\begin{table}[H]
\tiny
\caption{Group differences by content creatorship in the percentage of participants that reported aggregate dimensions in behaviors that reduce psychological safety in Communities of Practice, and the effect on participants (N=143). Groups with fewer than 3 participants are suppressed to protect anonymity. Statistical group differences ($p < .05$) are marked with *.}
\label{tab:results-phase1-groupdifferencesqualitative-contentcreatorship}
\begin{tabular}{lcccc}
\toprule
\textbf{Dimension} & \textbf{All} & \textbf{Yes} & \textbf{No} & \textbf{p} \\
\midrule
\multicolumn{5}{l}{\textbf{Behaviors contributing to low psychological safety}} \\
\midrule
Negative interactions & 33.8\% & 34.6\% & 32.7\% & .825 \\
Hostile behaviors & 33.1\% & 38.3\% & 24.5\% & .107 \\
Exclusionary behaviors & 63.1\% & 67.9\% & 55.1\% & .145 \\
Unsafe environment & 8.5\% & 9.9\% & 6.1\% & .460 \\
No experience & 9.2\% & 11.1\% & 6.1\% & .345 \\
\midrule
\multicolumn{5}{l}{\textbf{Effects of low psychological safety}} \\
\midrule
Emotional response & 16.9\% & 14.8\% & 20.4\% & .414 \\
Avoid interpersonal risks & 20.0\% & 19.8\% & 20.4\% & .929 \\
Personal reflections & 8.5\% & 7.4\% & 10.2\% & .582 \\
Reduce contributions & 58.5\% & 54.3\% & 65.3\% & .221 \\
Improve situation & 10.8\% & 9.9\% & 12.2\% & .676 \\
No effect & 11.5\% & 12.3\% & 10.2\% & .714 \\
\bottomrule
\end{tabular}
\end{table}

\begin{table}[H]
\tiny
\caption{Role differences in the percentage of participants that reported aggregate dimensions in behaviors that reduce psychological safety in Communities of Practice, and the effect on participants (N=143). Groups with fewer than 3 participants are suppressed to protect anonymity. Statistical group differences ($p < .05$) are marked with *.}
\label{tab:results-phase1-groupdifferencesqualitative-role}
\begin{tabular}{lccccccc}
\toprule
\textbf{Dimension} & \textbf{All} & \textbf{Scrum Master} & \textbf{Agile Coach} & \textbf{Trainer or facilitator} & \textbf{Consultant} & \textbf{Other} & \textbf{p} \\
\midrule
\multicolumn{8}{l}{\textbf{Behaviors contributing to low psychological safety}} \\
\midrule
Negative interactions & 33.8\% & 45.5\% & 30.4\% & 40.0\% & 21.1\% & 20.0\% & .106 \\
Hostile behaviors & 33.1\% & 27.3\% & 32.6\% & 46.7\% & 21.1\% & 40.0\% & .198 \\
Exclusionary behaviors & 63.1\% & 60.6\% & 63.0\% & 66.7\% & 52.6\% & 73.3\% & .725 \\
Unsafe environment & 8.5\% &  & 6.5\% &  & 21.1\% &  & .276 \\
No experience & 9.2\% & 15.2\% & 8.7\% &  &  &  & .668 \\
\midrule
\multicolumn{8}{l}{\textbf{Effects of low psychological safety}} \\
\midrule
Emotional response & 16.9\% & 24.2\% & 13.0\% & 26.7\% &  &  & .285 \\
Avoid interpersonal risks & 20.0\% & 12.1\% & 26.1\% &  & 36.8\% &  & .142 \\
Personal reflections & 8.5\% & 12.1\% & 8.7\% &  &  &  & .165 \\
Reduce contributions & 58.5\% & 51.5\% & 58.7\% & 66.7\% & 52.6\% & 73.3\% & .47 \\
Improve situation & 10.8\% & 21.2\% & 8.7\% &  &  &  & .079 \\
No effect & 11.5\% & 18.2\% & 8.7\% &  & 15.8\% &  & .505 \\
\bottomrule
\end{tabular}
\end{table}

\subsubsection{Factors contributing to lowered psychological safety in Communities of Practice}\
The previous sections addressed differences in psychological safety between online and face-to-face interactions and between subgroups. We now turn to the factors contributing to decreased psychological safety. Participants were asked to identify the factors that decrease psychological safety in their Communities of Practice. We used the Gioia methodology described in Section~\ref{sec:results-phase1-analysis-qualitative} to develop a conceptual model of first-order concepts, second-order themes, and third-order aggregate dimensions reported in Figure~\ref{fig:gioia-factors} in Appendix~\ref{sec:appendix}.

Four aggregate dimensions in the factors emerged from the qualitative coding. The most reported dimension was ``negative interaction patterns'' (53.8\%), consisting of the themes ``negative communication'' (33.1\%), ``bullying and aggression'' (23.1\%), ``lack of empathy'' (12.3\%) and ``personal factors'' (3.8\%). The second most-reported dimension was ``tribalism'' (52.3\%), representing the themes ``rigid views'' (43.8\%) and ``exclusion'' (18.5\%). Next, 23.8\% of cases reported codes related to the aggregate dimension ``community dynamics'', consisting of the themes ``competition'' (17.7\%), ``online dynamics'' (6.2\%), and ``lack of support'' (2.3\%). 5.4\% of respondents reported factors related to the aggregate dimension ``leadership and governance'', with the underlying themes ``lacking governance'' (3.8\%), ``lack of positive modelling'' (1.5\%) and ``leaders act badly'' (1.5\%).

In summary, participants broadly consider the style and quality of interactions to be the most important regarding psychological safety, closely followed by a community's openness to new ideas and participants.

\subsubsection{Strategies to improve psychological safety in Communities of Practice}\
In this section, we examine strategies to enhance psychological safety. Participants were asked to propose strategies that could improve psychological safety for themselves and others within their Community of Practice. We used the Gioia methodology described in Section~\ref{sec:results-phase1-analysis-qualitative} to develop a conceptual model of first-order concepts, second-order themes, and third-order aggregate dimensions reported in Figure~\ref{fig:gioia-strategies} in Appendix~\ref{sec:appendix}.. 

Five aggregate dimensions emerged from the qualitative coding of strategies. The most reported dimension consisted of codes related to ``improve interactions'' (33.1\%), consisting of the themes ``develop soft skills'' (14.6\%), ``change personal mindset'' (13.1\%), ``communication and interaction strategies'' (9.2\%) and ``build awareness on psychological safety'' (5.4\%). The second most-reported dimension was ``improve dynamics'' (30.0\%), representing the themes ``community norms'' (15.4\%), ``stand up to norm violations'' (6.9\%), improving ``diversity and inclusion'' (6.2\%) and ``structural changes'' (18.5\%). Next, 20.8\% of cases reported codes related to the aggregate dimension ``leadership and moderation'', consisting of the themes ``lead by example'' (13.1\%), ``moderation'' (6.9\%), and ``leadership style'' (1.5\%). 17.7\% of respondents reported factors related to the aggregate dimension ``personal coping'', with the underlying themes ``acceptance'' (2.3\%) and recognizing psychological safety as a ``personal responsibility'' (2.3\%). Finally, 17.7\% of the participants reported codes in the category ``do not know''.

In summary, improving psychological safety in Communities of Practice is not easy, as evidenced by the substantial percentage of participants who report not knowing how. However, most participants see the development of soft skills as an enabler for respectful and appropriate debate. This puts responsibility on both sides in interactions. Additionally, Communities of Practice can take action to create more positive dynamics among members by establishing explicit norms and holding individuals accountable to them.

\section{Phase~II: Member checking}
\label{sec:phase2}
The first phase of this study assessed the level of psychological safety in face-to-face and online interactions within Communities of Practice. It explored factors contributing to lower psychological safety and strategies to enhance it. Since psychological safety is inherently subjective, we sought to confirm our findings by engaging in member checking during the second phase of our study. Member checking is a research method where researchers share their findings with participants to assess credibility and collect feedback~\cite{lincoln1986research}. Given the collaborative focus of our research, this approach also enabled us to draw on participants’ unique perspectives in interpreting the results and co-generating new research questions. Accordingly, we adopted the member checking strategy described by McKim~\cite{mckim2023meaningful}.

\subsection{Participants}
\label{sec:phase2participants}
We conducted our data collection process for Phase~II through an online survey deployed using IBM Qualtrics. Data collection was performed between June 2024 and July 2024. Like Phase~I, a purposive non-probabilistic sampling strategy~\cite{baltes2022sampling} was pursued. Due to the anonymous data collection in Phase~I, prior participants could not be invited personally. Instead, the survey was promoted in the same Agile Communities of Practice as Phase~I. A question was included in the survey to determine if participants had participated in Phase~I.

The survey consisted of Likert scales to capture quantitative data and essay questions to collect qualitative data, as described in Section~\ref{sec:phase2measurements}. The complete survey for Phase~II is included in Table~\ref{tab:appendix-phase2-survey} in Appendix~\ref{sec:appendix}.

\begin{figure}[H]
\centering
\includegraphics[width=3in]{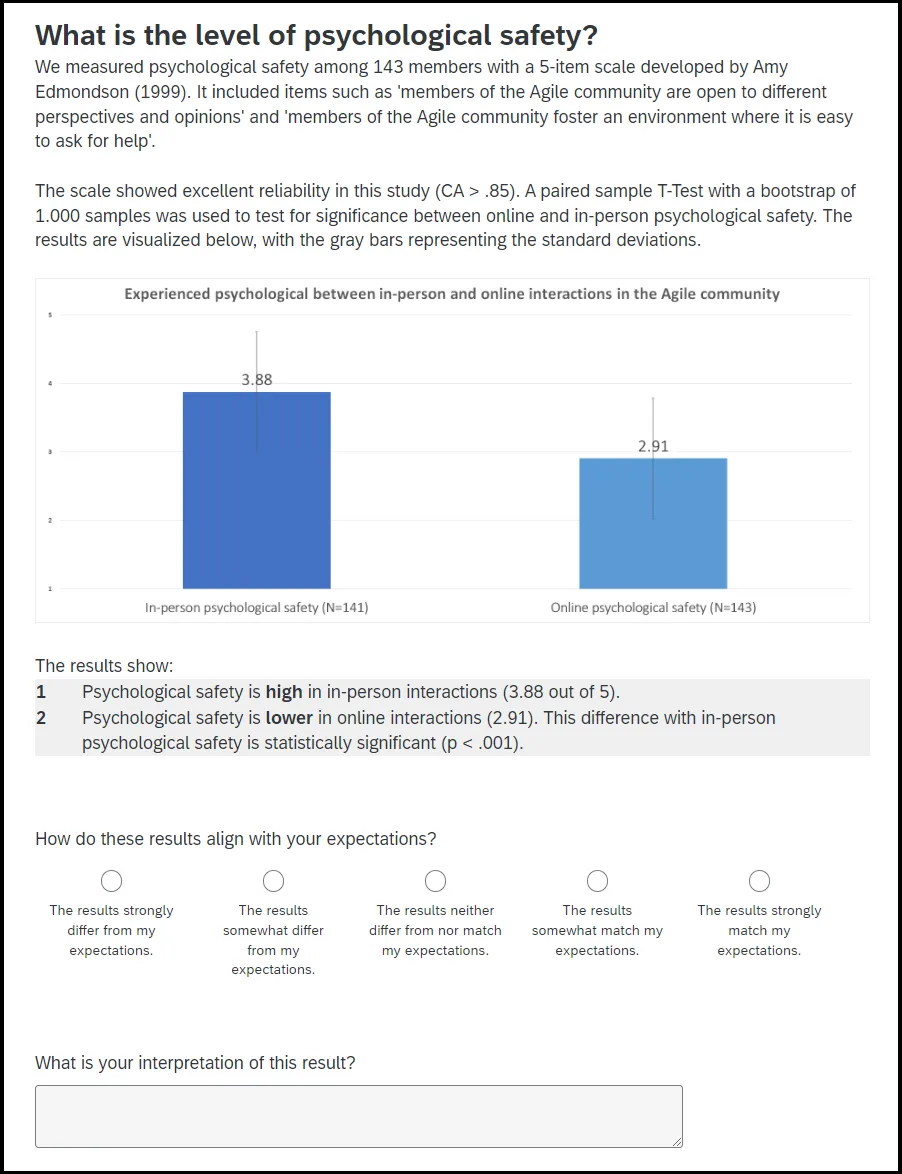}
\caption{Example of one step in the survey for Phase~II}
\label{fig:phase2surveyexample}
\Description[Screenshot of the survey in Phase~II]{The page shows how we present results to participants in Phase~II. First, a short description of the analysis method is provided, followed by a chart. Below the chart are key findings.}
\end{figure}

Member checking typically involves sharing raw responses and the inferences made from them with participants. This requires significant time from participants when a substantial dataset is involved, resulting in low response rates and limited feedback~\cite{stake1995art,mckim2023meaningful}. Therefore, we employed a more structured approach outlined by McKim~\cite{mckim2023meaningful}. The survey presented the results from Phase~I through visualizations (i.e., histograms and bar charts), as well as core findings per topic. The core findings were presented as factually as possible to reduce bias (i.e., ``Psychological safety is high in face-to-face interactions (3.88 out of 5)'' and ``Psychological safety is low in virtual interactions (2.91 out of 5)''). Participants were asked to rate the extent to which the findings matched their expectations and to provide interpretations through essay questions. An example is shown in Figure~\ref{fig:phase2surveyexample}.

A total of 31 participants completed the survey. Since the survey was public, a proper response rate calculation is impossible. A data review yielded one careless response with mostly empty answers~\cite{meade2012identifying}, which was removed. The sample composition ($N=30$) is reported in Table~\ref{tab:phase2samplecomposition}.

\begin{table}[H]
\centering
\caption{Composition of the sample for Phase~II}
\label{tab:phase2samplecomposition}
\begin{tabular}{m{4cm}p{3cm}p{1.4cm}}
\toprule
\textbf{Variable} & \textbf{Category} & \textbf{N (\%)} \\ \midrule
Respondents &  & 30 (100\%) \\
Participated in Phase~I & Yes & 8 (26.7\%) \\
 & No & 15 (50\%) \\ 
 & Do not remember & 7 (23.3\%) \\ 
 \bottomrule
\end{tabular}
\end{table}

\subsection{Measurements}
\label{sec:phase2measurements}
\textbf{Alignment of results with expectations by participants}: The results from the quantitative measures in Phase~I were presented divided across discrete steps in the survey, each with a 5-point Likert item to assess the degree to which these results matched the expectations of the participant (``How do these results align with your expectations?''). Detailed results from the qualitative analyses in Phase~I were omitted to avoid overloading participants. Alignment was evaluated for four areas of the results: psychological safety between online and face-to-face interactions, gender differences, role differences, and age differences. Seniority and content creatorship were omitted for brevity.

\textbf{Interpretation of results:} To collect interpretations from participants, one essay question was included in each step of the survey that presented results: ``What is your interpretation of this result?''. Additionally, one essay question was included at the end of the survey to collect follow-up questions: ``Based on the results you've interpreted so far, what follow-up questions come to mind for you?''. This qualitative data was coded as described in Section~\ref{sec:phase2-analysis}.

\textbf{Interest in research process}: This was assessed with a 5-point Likert item (``To our knowledge, this is the first time the community actively contributes to the sense-making of scientific evidence. Did you find this process interesting?''). Figure 24 shows that the mean average varied between 4.13 and 4.38, reflecting answers between “Very interesting” and “Extremely interesting”.

\textbf{Participation in Phase~I}: Near the end of the survey, participants were asked if they remembered participating in Phase~I. A single categorical item was used - ``Did you participate in the survey on psychological safety yourself?'' - with the options ``Yes'', ``No'', and ``Do not remember''.

\subsection{Analysis}
\label{sec:phase2-analysis}
The quantitative data was analyzed using IBM SPSS. Outlier detection was performed with stem-and-leaf plots. While some outliers were present on both sides of the distribution, we retained them as they represented genuine cases. No data was missing, so all records were included in the analyses. The distributions of our continuous measures were significantly non-normal, as indicated by Kolmogorov-Smirnov tests, although kurtosis ($<3$) and skewness ($<2$) remained below their recommended thresholds~\cite{hair2019multivariate}. The assumption of equal variance was not violated, as Levene's test returned a non-significant result for each of our continuous measures. To reduce bias in parameter estimates due to non-normality, a nonparametric Kruskal-Wallis independent samples median test was applied to compare medians across subgroups based on prior participation in Phase~I.

The first author coded qualitative responses into broad themes.

We performed a post hoc power analysis using G*Power~\cite{faul2009statistical}, version 3.1.9. We determined that our sample for Phase~II allows us to capture small effects ($\textit{d}=.020$) with a power of ~60\% ($1-\beta= .600$) and 87.4\% ($1-\beta= .874$) for moderate effects ($\textit{d}=.050$). Thus, we are confident that we can detect most moderate effects.

\subsection{Results}
The results for the quantitative measures are reported in Table~\ref{tab:results-phase2}. No significant differences were found based on whether participants also participated in Phase~I, indicating no bias due to prior participation.

\begin{table}[!ht]
\centering
\tiny
\caption{Alignment of results with expectations for participants in Phase~II. Means, Standard deviations, significance for all participants in Phase~II based on non-parametric Kruskal-Wallis ANOVA (N=30), participants who also participated in Phase~I (N=8), participants who did not participate in Phase~II (N=15), and participants for whom it is unknown if they participated in Phase~I (N=7).}
\label{tab:results-phase2}
\begin{tabular}{lccccccccc}
\toprule
\textbf{Variable} & \multicolumn{2}{c}{\textbf{All (N=30)}} & \multicolumn{2}{c}{\textbf{In phase 1 (N=8)}} & \multicolumn{2}{c}{\textbf{Not in phase 1 (N=15)}} & \multicolumn{2}{c}{\textbf{Unknown (N=7)}} & \multicolumn{1}{c}{\textbf{Group difference}} \\
\midrule
\textbf{} & \textit{M} & \textit{SD} & \textit{M} & \textit{SD} & \textit{M} & \textit{SD} & \textit{M} & \textit{SD} & \textit{p} \\
\multicolumn{10}{c}{\textit{Alignment with participant expectations}} \\
Psychological safety face-to-face vs online & 4.167 & .913 & 4.375 & 1.061 & 4.000 & .926 & 4.286 & .756 & .175 \\
Psychological safety by gender & 3.933 & .828 & 3.875 & .991 & 3.800 & .775 & 4.286 & .756 & .698 \\
Psychological safety by role & 3.333 & 1.061 & 3.375 & 1.061 & 3.267 & 1.100 & 3.429 & 1.134 & .587 \\
Psychological safety by age & 3.067 & 1.258 & 3.000 & .926 & 3.000 & 1.464 & 3.286 & 1.254 & .310 \\
\multicolumn{10}{c}{\textit{Evaluation}} \\
Interest in research process & 4.233 & .679 & 4.375 & .744 & 4.133 & .640 & 4.286 & .756 & .407 \\
\bottomrule
\end{tabular}
\vspace{4em}
\end{table}

\subsubsection{Evaluation by participants of difference between face-to-face and online interactions}
The level of psychological safety reported in Phase~I, as well as the difference between online and face-to-face interactions, matched the expectations of most participants ($M=4.167, SD=.913, N=30$). One participant noted ``I feel safer to be open and honest when meeting in person vs online''. Another wrote ``This does not surprise me, I suspect that you see this difference between in-person and social media engagement in general''. However, some participants expected larger differences: ``I would expect in-person to score higher and online to score a lot lower''. Table~\ref{tab:hypotheses-differenceonlineandfacetoface} reports four Hypotheses generated by participants. Seven participants suggested that the lack of social cues and body language reduces the potential for psychological safety in online interactions (i.e. \textit{``in-person interactions provide more social cues to support understanding''}). Seven participants hypothesized that online interactions typically involve more people who are unfamiliar with each other, and those weaker social bonds limit psychological safety (i.e. \textit{``I assume that with local meetups people already have a relationship with their peers.}). Third, participants offered that people in online interactions are less willing or capable of showing empathy, making their responses less safe (i.e. \textit{``It seems people naturally care less about being polite and thoughtful when disagreeing [in online interactions]''}). Finally, one participant pointed out that online interactions may be more ad-hoc and less facilitated, with fewer opportunities to build psychological safety.

\begin{table}[!ht]
\centering
\tiny
\caption{Hypotheses offered by participants to explain the reported difference in psychological safety for face-to-face and online interactions ($N=28$). Note that participants can be coded into one or more categories.}
\begin{tabular}{lc}
\toprule
\textbf{Hypothesis} & \textbf{Number of participants} \\
\midrule
Lack of social cues and body language in online interactions makes it harder to build psychological safety & 7 \\
In online interactions, people have weaker social bonds or personal connections & 7 \\
People are less capable or willing to show empathy in online interactions & 6 \\
Online interactions are less prepared and facilitated & 1 \\
Other & 7 \\
\bottomrule
\end{tabular}
\vspace{4em}
\label{tab:hypotheses-differenceonlineandfacetoface}
\end{table}

\subsubsection{Evaluation by participants of gender difference for psychological safety}
Gender differences were mainly as expected by participants for overall psychological safety ($M=3.933, SD=.828, N=30$). The participants generated four discrete hypotheses, listed in Table~\ref{tab:hypotheses-genderdifferences}. Ten participants ventured that the results reflected the gender bias and male dominance in communities at large (i.e. ``\textit{Matches both my experience in society and in online interactions per se as a woman}''). Four participants attributed the differences to a lower sensitivity to their psychological safety and that of others in men compared to other genders (i.e. \textit{``Men may not be aware that those of other genders feel uncomfortable speaking up and are more likely to feel safe themselves.''}). A third, related hypothesis offered by two participants was that men are typically more self-confident, which makes them less sensitive. Finally, one participant noted that gender differences may be less salient in online interactions, which explains the small difference reported specifically for online interactions. 

\begin{table}[!ht]
\centering
\tiny
\caption{Hypotheses offered by participants to explain the reported gender difference in psychological safety for face-to-face and online interactions ($N=24$). Note that participants can be coded into one or more categories.}
\begin{tabular}{lc}
\toprule
\textbf{Hypothesis} & \textbf{Number of participants} \\
\midrule
Gender bias and male dominance in communities at large & 10 \\
Men are less sensitive to situations of low psychological safety & 4 \\
Men are more self-confident than other genders & 2 \\
Gender is less visible in online interactions & 1 \\
Other & 7 \\
\bottomrule
\end{tabular}
\vspace{4em}
\label{tab:hypotheses-genderdifferences}
\end{table}

\subsubsection{Evaluation by participants of role difference for psychological safety}
The differences in psychological safety by role ($M=3.333, SD=1.061, N=30$) aligned less with expectations. Participants broadly generated four discrete Hypotheses, listed in Table~\ref{tab:hypotheses-differenceonlineandfacetoface}. Nine participants speculated that roles vary in terms of personal stake in creating psychological safety, and are thus more sensitive to it (i.e. ``\textit{As a Scrum Master, I often talk about psychological safety with my teams and others around my team. Will that make me more likely to feel safe myself [...]?}''). Five participants hypothesized that some roles invite more criticism, such as trainers and consultants, who are typically seen as bringing solutions. Another potential factor is that some roles typically spend more time with the same people, and are thus more likely to establish psychological safety (i.e. ``\textit{Scrum Masters may work with the same groups more frequently than trainers}''). Finally, two participants speculated that higher needs for ego confirmation accompany some roles (``\textit{Perhaps it is a challenge to the ego of trainer/facilitator.}''). 

\begin{table}[!ht]
\centering
\tiny
\caption{Hypotheses offered by participants to explain the reported role difference in psychological safety for face-to-face and online interactions ($N=26$). Note that participants can be coded into one or more categories.}
\begin{tabular}{lc}
\toprule
\textbf{Hypothesis} & \textbf{Number of participants} \\
\midrule
Level of personal interest in creating psychological safety varies by role & 9 \\
Proneness to criticism due to role (i.e. trainers, consultants) & 5 \\
Certain roles spend more time with groups, thus allowing for more safety to form & 2 \\
Certain roles may be accompanied by higher need for ego confirmation (i.e. trainer / facilitator) & 2 \\
Other & 8 \\
\bottomrule
\end{tabular}
\vspace{4em}
\label{tab:hypotheses-roledifferences}
\end{table}

\subsubsection{Evaluation by participants of age difference for psychological safety}
The differences in psychological safety by age ($M=3.067, SD=1.258, N=30$) were least similar to what participants expected. Three discrete hypotheses were extracted from the responses, listed in Table~\ref{tab:hypotheses-agedifferences}. Eleven participants speculated that generational differences make some people more sensitive to their psychological safety and/or that of others (i.e. ``\textit{is it so that the older you get, the safer you feel due to your level of maturity?}''). Related to this, five participants speculated that different generations have different levels of experience with online interactions, i.e. ``\textit{Younger groups [are] very much used to not taking everything serious on social media, [whereas] oldest people [are] leaning on their experience. Middle-aged [are] having the most troubles?}''. Finally, one participant noted that the reported effects may be due to younger generations having different roles, and different levels of exposure to psychological (un)safety.

\begin{table}[!ht]
\centering
\tiny
\caption{Hypotheses offered by participants to explain the reported age difference in psychological safety for face-to-face and online interactions ($N=26$). Note that participants can be coded into one or more categories.}
\begin{tabular}{lc}
\toprule
\textbf{Hypothesis} & \textbf{Number of participants} \\
\midrule
Generational differences in sensitivity to safety and unsafety & 11 \\
Generational differences in how we interact online shape how we respond to it & 5 \\
Younger age groups are correlated with other types of roles & 1 \\
Other & 9 \\
\bottomrule
\end{tabular}
\vspace{4em}
\label{tab:hypotheses-agedifferences}
\end{table}

\subsubsection{Follow-up research questions} 
At the end of the survey, participants were also asked to generate more follow-up questions based on the results. Five discrete themes were extracted from the responses, listed in Table~\ref{tab:hypotheses-followupquestions}. Most participants were curious about how the results compared to other or larger populations (i.e. ``I feel that the problems we have in our community exist on a broader scale, so curious what we could do about this in such a large group''). Several participants also made suggestions to improve Phase~I, particularly some definitions related to Agile terminology. Three participants also wondered how to improve psychological safety, as well as the specific factors contributing to it. Other participants were interested in different ways to measure psychological safety and how it correlated with (team) performance (i.e. ``Do you know the connection between psychological safety and performance?''). Finally, two participants suggested exploring certain factors, specifically social factors, and how the type of interaction shapes psychological safety.

\begin{table}[!ht]
\centering
\tiny
\caption{Themes in follow-up questions offered by participants based on the results from Phase~I ($N=18$). Note that participants can be coded into one or more categories.}
\begin{tabular}{lc}
\toprule
\textbf{Hypothesis} & \textbf{Number of participants} \\
\midrule
Comparison to other or more general populations & 4 \\
Inquiries or feedback on Phase~I & 4 \\
How to improve safety (i.e. how to decrease dogmatism, how to improve online, role of leadership) & 3 \\
Other measures of psychological safety and correlations with outcomes (i.e. performance) & 2 \\
Investigate role of potential factors (i.e. kind of interaction, social factors) & 2 \\
Other & 6 \\
\bottomrule
\end{tabular}
\vspace{4em}
\label{tab:hypotheses-followupquestions}
\end{table}

\subsubsection{Interest in the research process}
Finally, participants were asked to rate how interesting they found the process of reviewing Phase~I results and generating ideas. Interest was high ($M=4.233, SD=.679, N=30$). Participants described how reviewing the results was useful, with quotes such as ``\textit{It's fun and helps with self-reflection}'', ``\textit{Very interesting - please continue}'' and ``\textit{Makes me feel valued and valuable}''. One participant was more skeptical and wrote ``\textit{Everyone will respond to surveys differently based on their context. I do not feel it really rendered realistic results. Despite it being anonymous}''. Several participants noted that while the process was valuable and engaging, the amount of data was quite overwhelming and could have been grouped more effectively. Two participants also noticed two mistakes in the visualizations in the results.

\section{Discussion}
\label{sec:Discussion}
This study examined psychological safety in online and face-to-face interactions within Agile Communities of Practice. Using a multi-phased, mixed-method design and member checking, we involved community members in both data collection and interpretation. In Phase I, 143 participants reported significantly lower psychological safety in virtual settings, with age and role, but not gender, seniority, or content creation, modulating these perceptions. Exclusionary behaviors and negative interactions emerged as primary threats, often resulting in reduced contributions and emotional responses in participants. Improvement strategies centered on developing soft skills, establishing explicit norms, and ensuring supportive leadership. In Phase II, participants validated the findings and offered constructive reflections, confirming their relevance and credibility.

\subsection{Factors Contributing to Low Psychological Safety}

Communities of Practice (CoPs) support learning through sustained peer interaction~\cite{wenger2002cultivating}. As described in Legitimate Peripheral Participation Theory~\cite{lave1991situated,wenger2009communities}, members typically move from peripheral involvement to full participation. Our findings suggest that this progression can halt or reverse when psychological safety is compromised, leading to disengagement across demographic lines. This outcome reduces knowledge-sharing intentions~\cite{chen2007factors} and disrupts reciprocity~\cite{hsu2007knowledge,wasko2000one,chiu2006understanding}, thereby weakening the social learning process that underpins CoPs. The significantly lower levels of psychological safety in virtual contexts further highlight the urgency of addressing these issues in digitally mediated communities.

While these results align with prior literature, our qualitative data offer additional insight into underlying causes. The most frequently reported contributor was negative communication (53.8\%), which included both overt behaviors (e.g., bullying, personal attacks) and more subtle ones, such as sarcasm or inappropriate humor. Such ambiguity is heightened in online settings where non-verbal cues are absent~\cite{kostovich2020establishing,hayashida2024privacy}, consistent with prior findings on the challenges of digital collaboration~\cite{storey2014r}.

Improving communication quality is thus central. While technological solutions such as emoticons attempt to convey tone, they can increase ambiguity~\cite{edwards2017s,siegel1986group}. More effective are interventions targeting soft skills that help participants understand how their behavior affects others. Our findings provide a receiver-centric view of these dynamics, complementing O’Donovan et al.’s~\cite{o2020measuring} broader observational taxonomy of behaviors, including voice, silence, supportiveness, and familiarity. Increasing awareness of such categories may help individuals communicate with greater psychological sensitivity.

Tribalism also played a prominent role, particularly in the form of rigid group boundaries and exclusionary practices. Social Identity Theory~\cite{tajfel1979integrative} helps explain this tendency to prioritize in-group cohesion, which may be exacerbated by group polarization~\cite{lord1979biased,isenberg1986group}, self-censorship, and groupthink~\cite{park1990review}. Although such tendencies are evolutionarily rooted~\cite{geary2005origin}, they can be moderated. Strategies include establishing shared norms~\cite{cave2016centre}, articulating clear goals~\cite{preece2001sociability}, and modeling inclusive behavior through leadership~\cite{lechner2022create}.

Structurally, digital platforms can support psychological safety through moderation and intentional design that fosters reflection. Facilitation methods such as Liberating Structures~\cite{lipmanowicz2015liberating,kimball2012liberating}—also cited by participants—offer practical scaffolding by providing structured interaction formats~\cite{kimball2012liberating}. Although empirical studies are still limited, early evidence suggests they may enhance inclusion and empathy in group learning~\cite{singhal2020liberating,ferguson2015applying}.

\subsection{Awareness of Psychological Safety as a Moderator}

As anticipated, \textbf{few differences in psychological safety were observed across demographic variables}. In line with Zhang et al.~\cite{zhang2010exploring}, no significant gender differences emerged. However, qualitative data indicated that men reported fewer negative interactions than other genders — 25\% compared to approximately 50\%. As noted in Phase II, this may reflect gendered norms in both expressing and perceiving psychological safety, and aligns with findings that women more frequently engage in behaviors that foster it~\cite{heijdens2022role}.

In contrast to Zhang et al.~\cite{zhang2010exploring}, \textbf{we found notable age-related differences in online settings}. Participants aged 36–55 reported the lowest levels of psychological safety. Responses suggest a generational effect: younger participants (26–35) may be more attuned to the tone and tempo of online interactions, while older participants (56+) may be more detached but less affected. Those in between—often digital migrants—may experience the greatest friction as they navigate between offline norms and online expectations~\cite{azad2023different,mude2023social}.

Role-based patterns mirror this dynamic. Scrum masters, often in younger or mid-level roles, reported the highest psychological safety, whereas consultants and trainers—typically older and more senior—reported the lowest. A post-hoc analysis confirmed a significant association between age and role ($Cramer's V = .274$, $p < .001$). These variations may reflect differing degrees of immersion in team dynamics, as well as differing professional stakes in cultivating psychological safety~\cite{shastri2021spearheading}.

\textbf{These findings suggest that psychological safety may be moderated by awareness and role-based sensitivity.} Individual differences—shaped by age, role, gender, or professional experience—likely influence both how safety is created and how it is perceived. Yet, heightened sensitivity can also lead to misinterpretation. As one participant noted, “cancel culture and general political culture make it nearly impossible to have an in-depth conversation.” Others viewed psychological safety as a personal responsibility and raised concerns about its impact on freedom of expression.

Such reactions point to a common misconception: that psychological safety aims to eliminate conflict. The opposite is true. As Edmondson~\cite{edmondson2014psychological} notes, psychological safety enables disagreement by ensuring that it does not carry social cost. Constructive conflict, openly expressed and respectfully handled without permanently harming relationships, is a sign of psychological safety, not its absence. Ultimately, it is not sensitivity to discomfort that matters most, but sensitivity to the conditions that make open dialogue possible.

\subsection{Implications for Practice}
This study highlights several actionable insights for those designing, leading, or facilitating Communities of Practice (CoPs). Psychological safety directly shapes whether members feel comfortable contributing, especially in virtual settings where cues are limited and misinterpretations more likely~\cite{zhang2010exploring}. Low safety hinders centripetal participation~\cite{lave1991situated}, thereby weakening CoPs as social learning structures.

\textbf{First, psychological safety must be made an explicit, shared priority.} Leaders and veteran members should not assume others feel equally secure. Behaviors such as sarcasm or blunt feedback may be routine for some, yet alienating to others, depending on culture or history. Psychological safety should be openly discussed using concrete examples. This study’s conceptual model, Edmondson’s foundational work~\cite{edmondson1999psychological,edmondson2006explaining,edmondson2023psychological}, and frameworks such as O’Donovan et al.~\cite{o2020measuring} offer helpful starting points. Communities should translate these into co-created norms—e.g., on disagreement, welcoming newcomers, or interpreting silence—but ensure they are consistently modeled~\cite{lechner2022create}. Still, the efficacy of such awareness initiatives remains mixed~\cite{scarpinella2023can,campbell2024exploring,dusenberry2020building}.

\begin{figure}
\centering
\includegraphics[width=1.0\linewidth]{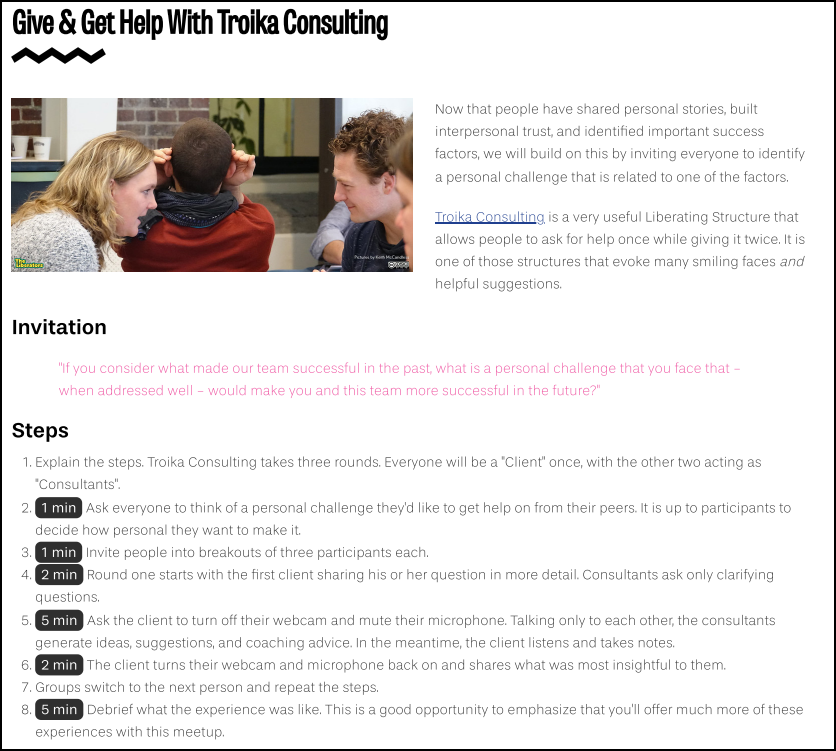}
\caption{Facilitation instructions for the Liberating Structure ``Troika Consulting'' used in a virtual Community of Practice by the first author.}
\label{fig:troika}
\end{figure}

\textbf{Second, structure interactions to support inclusion.} Liberating Structures~\cite{lipmanowicz2015liberating}, such as “1-2-4-All” or “Troika Consulting,” can decentralize participation and create room for quieter voices. In virtual settings, practices like rotating facilitators or anonymous input channels can help mitigate dominance effects. Figure~\ref{fig:troika} shows how “Troika Consulting” was implemented virtually to foster equal participation.

\textbf{Third, address tribalism directly.} Participants frequently noted subtle exclusionary behaviors. Social Identity Theory~\cite{tajfel1979integrative} helps explain these dynamics, but they undermine learning and inclusion. Leaders can mitigate them by modeling openness, rotating roles, or pairing newcomers with experienced members. Even small actions signal that diverse perspectives are welcome~\cite{wenger2002cultivating}.

\textbf{Finally, experiment locally.} Most interventions remain generic or untested. Participatory approaches, such as action research, allow CoPs to test practices suited to their own context. Suggestions like visual tone indicators or reflective check-ins can yield high-impact, low-cost changes when co-developed and evaluated in situ.

\subsection{Limitations}
This section outlines the limitations of our study, guided by Russo et al.~\cite{Russo2018ISQ} and incorporating both qualitative and quantitative considerations. We address threats to credibility, transferability, dependability, and confirmability~\cite{guba1981criteria}, along with internal, construct, external, and conclusion validity~\cite{wohlin2012experimentation}. To promote transparency and reproducibility, we have made anonymized quantitative data, syntax files, and codebooks openly available on Zenodo\footnote{A replication package is openly available under a CC-BY-NC-SA 4.0 license on Zenodo, DOI: \url{https://doi.org/10.5281/zenodo.15767147}.}.

\textbf{Credibility \& Internal}. 
In qualitative research, credibility refers to the degree to which the findings accurately reflect the perspectives and experiences of participants~\cite{lincoln1985naturalistic}. We employed the Gioia framework~\cite{gioia2012organizational} to rigorously structure and code the data through open and axial coding as part of our thematic analysis. Dual coding by two researchers yielded high inter-rater agreement ($\kappa > .90$), reducing interpretative bias. Additionally, the process was overseen by an expert researcher. See Section~\ref{sec:results-phase1-analysis-qualitative} for details and the replication package for the full codebook. Member checking was incorporated in Phase~II, which confirmed broad recognition of the findings by participants, thereby reinforcing credibility~\cite{lincoln1986research}.

Internal validity concerns the extent to which observed effects can be causally attributed~\cite{cook1979quasi}. We confirmed statistical assumptions~\cite{hair2019multivariate} and applied robust methods, including bootstrapping, Welch’s ANOVAs (Phase~I), and Kruskal-Wallis median tests (Phase~II), to mitigate distributional violations. Triangulation across qualitative and quantitative data sources further strengthened the robustness of inferences. While self-selection bias is a common limitation in voluntary surveys, we addressed this through anonymized participation, inclusion of opt-out options, and broad dissemination (Section~\ref{sec:phase1}). The variation in psychological safety scores, including 9.2\% of respondents reporting no negative experiences, suggests sample diversity.

\textbf{Transferability \& External}.
Transferability and external validity concern the extent to which results can be generalized beyond the study sample~\cite{goodwin2016research}.  Two limitations apply. First, our sample consisted of members from Communities of Practice in Agile Software Development. The results may not necessarily transfer to other domains. Second, due to purposive, non-probabilistic sampling, our sample may not represent all Communities of Practice in this domain. 

We addressed this in three ways. First, we provide detailed sample descriptions in Section~\ref{sec:phase2participants} to enable researchers to assess the transferability of the findings to their settings. Second, the community-level validation performed in Phase~II demonstrated that the results aligned substantially with those of the participants. No significant difference was found between participants who had participated in Phase~I and those who did not. Third, we promoted the study across various platforms, communities, and media, as described in Section~\ref{sec:phase1}, and contacted community leaders to help spread awareness.

\textbf{Dependability \& Construct}. 
Dependability in qualitative research refers to the consistency and reliability of findings throughout the research process~\cite{goodwin2016research}. Qualitative data was translated into insights through the Gioia methodology~\cite{gioia2012organizational}, which aims for scientific rigor in qualitative analyses. Two coders independently coded all cases for Phase~I to reduce bias and iterated on a codebook through memoing. Inter-rater reliability indicated high agreement ($\kappa > .90$). The codebook and memos are available in the replication package.

Construct validity refers to the degree to which the measures used in quantitative analyses measure their intended constructs~\cite{cook1979quasi}. Psychological safety was measured with an established scale developed and validated by Edmondson~\cite{edmondson1999psychological}, once for face-to-face and once for online interactions. A confirmatory factor analysis (CFA) confirmed that both measures represented distinct factors (see Table~\ref{tab:appendix-phase1-cfa} in Appendix~\ref{sec:appendix}). The reliability for both measures exceeded the cutoff recommended in the literature ($CR>=.70$~\cite{hair2019multivariate}). Thus, we are confident that we reliably measured the intended constructs.

\textbf{Confirmability \& Conclusion validity}. 
In qualitative research, confirmability is the extent to which the findings are grounded in the data and not the researcher’s biases or preferences~\cite{goodwin2016research}. To establish confirmability, we transparently documented our analysis process in this paper to allow traceability and auditing of our interpretations. We utilized open-source tools (QualCoder~\cite{curtain2025qualcoder}) for our coding and provided the codebook and memos in the replication package.

Conclusion validity refers to the soundness of inferences about relationships among variables~\cite{cozby2012methods}. The Phase~I sample was powered to detect moderate effects ($d = .50$) with 99.8\% confidence. We applied appropriate statistical techniques per recommended guidelines~\cite{hair2019multivariate} and triangulated data sources to ensure coherence of findings (Sections~\ref{sec:results-phase1-analysis-qualitative}, ~\ref{sec:results-phase1-analysis-quantitative} and~\ref{sec:phase2-analysis}).

\hyphenation{experience}
\begin{table}[!ht]
\centering
\tiny
\caption{Summary of Findings and Implications}
\label{tab:implications}
\begin{tabular}{p{2cm}p{5cm}p{5cm}}
\toprule
 & \textbf{Findings} & \textbf{Implications} \\ \midrule
\textbf{Online interactions are less safe}
& Psychological safety was lower in online interactions ($M=2.910, SD=1.027, N=141$) than face-to-face interactions ($M=3.879, SD=.880, N=141$).
& Due to the lack of verbal and nonverbal cues, psychological safety is harder to establish in online interactions. CoP organizers must carefully consider tools and practices to build safety and engage the community in creating behavioral norms for how to interact. \\ \addlinespace
\textbf{Themes for psychological safety}
& Based on qualitative responses from CoP members ($N=120$), we identified themes related to psychological safety in Communities of Practice using the Gioia methodology. Exclusionary behaviors, negative interaction patterns, and hostile behavior primarily lowered psychological safety. The most reported effects of low safety were reduced contributions, avoidance of interpersonal risks, and emotional responses. The factors that decreased psychological safety most were negative interaction patterns, tribalism, and community dynamics. 
& The results highlight the diversity of behaviors that decrease safety. Some behaviors have a clear negative intent, but most are more subjective and ambiguous. While one could argue that interpretation is a personal responsibility of the receiver, this ignores the reality that if many people feel unsafe, their withdrawal will undermine the learning purposes of a CoP and also impact the sender, thus making it a shared responsibility.
\\ \addlinespace
\textbf{Demographic differences for age and role, but not gender, seniority, and content creatorship}
& No significant differences were found in psychological safety between genders, levels of seniority, and content creatorship for online and face-to-face interactions. However, a significant difference was found between age groups for online interactions ($p < .01, \eta^2 = .138, 90\% CI [.025, .182]$) and between roles for face-to-face interactions ($p < .031, \eta^2 = .094, 90\% CI [.000, .138]$). Due to a correlation between cohort and age group, this may indicate a generational effect or role-related differences in awareness of psychological safety.
& While psychological safety is experienced similarly across demographic groups, some may be more sensitive or aware of psychological safety. What constitutes safe behavior for one person may not for another. Moreover, some subgroups may be targeted more by unsafe behaviors than others. This highlights both the need for and the complexity of diversity in groups. Organizers of CoPs have to be mindful of such differences and consider them in community guidelines. \\ \addlinespace
\textbf{Interventions to improve psychological safety}
& Based on qualitative responses by CoP members ($N=111$), we developed a conceptual model for strategies to improve psychological safety in Communities of Practice with the Gioia methodology. The most impactful strategies for improving psychological safety in CoPs are improving interactions through soft skills and interaction strategies, improving dynamics through structure and norms, and leadership.
& An underlying theme in most strategies is that people have different levels of awareness and sensitivity to psychological safety. The soft skills that typically create more effective interactions, like listening, asking questions, and showing appreciation, are relevant in CoPs too, particularly in online interactions. But this also takes time and effort, which is not always available. Thus, organizers of CoPs do well to invest in tools and facilitation techniques to encourage using such skills. \\ \addlinespace
\addlinespace
 \\ \bottomrule
\end{tabular}
\end{table}

\section{Conclusion}
\label{sec:Conclusion}

This study examined psychological safety in both virtual and face-to-face interactions within professional Communities of Practice (CoPs), with a focus on Agile Software Development. Using a two-phase, mixed-method design with member checking, we identified how psychological safety influences participation, and which contextual factors contribute to unsafe experiences.

Psychological safety was significantly lower in online interactions. While gender, seniority, and content creatorship showed no systematic effects, differences emerged by age and professional role: participants aged 36–55 and those in consulting or training roles reported less safety. These patterns suggest role- and generation-related sensitivities in digital settings.

Thematic analysis revealed that exclusionary behaviors, negative interaction patterns, and hostility are primary threats to safety. These behaviors reduce engagement, foster self-censorship, and contribute to emotional strain, weakening the capacity of CoPs to support sustained learning. Practitioner-suggested improvements emphasized four core levers: (i) developing soft-skill practices, (ii) co-creating behavioral norms, (iii) applying inclusive facilitation methods, and (iv) securing visible support from leadership. These strategies offer guidance to community organizers and platform providers aiming to foster more inclusive learning environments, particularly in digital contexts.

This study also advances research in three key ways: it is the first to compare psychological safety between face-to-face and virtual CoP interactions; it introduces an empirically grounded model linking unsafe behaviors, perceived consequences, and contextual drivers; and it incorporates practitioner input through member checking to align academic and practitioner priorities.

Future work should prioritize replication in domains beyond Agile Software Development to test the generalizability of these findings. Given possible domain-specific influences—such as community maturity or gender imbalance—comparative studies across fields with different cultures and compositions are warranted. Furthermore, there is a clear need for rigorous development and evaluation of targeted interventions, particularly those co-created with practitioners. Liberating Structures and other facilitative tools hold promise but require more empirical testing. Lastly, future studies should refine their treatment of interaction modalities. Greater granularity in assessing synchronous versus asynchronous, text versus video, or one-to-one versus one-to-many communication may clarify how these differences shape psychological safety. Complementing self-report data with behavioral and observational measures would strengthen causal inferences and support more robust theory-building.

\section{Acknowledgments}
\label{sec:Ack}
The authors would like to thank all participants and Communities of Practices participating in Phase~I and/or Phase~II of our study.

\section{Supplementary Materials}
\label{sec:Supplement}
A replication package for the study is available at the following DOI: 10.5281/zenodo.15767147 under a CC-BY-NC-SA 4.0 license. The package includes SPSS syntaxes, a codebook, and a coded dataset. Qualitative responses were omitted to protect the anonymity of participants.

\section{Responsible disclosure}
\label{sec:Disclosure}
The authors declare that they have no known competing financial interests or personal relationships that could have influenced the work reported in this paper.

\bibliographystyle{ACM-Reference-Format}
\bibliography{bib}

\newpage
\begin{appendix}
\section{Appendix}
\label{sec:appendix}

\begin{figure}[H]
\centering
\caption{First-order concepts, second-order themes, and aggregate dimensions in the effects of low psychological safety as reported in Communities of Practice by participants ($N=122$)}
\includegraphics[width=3.5in]{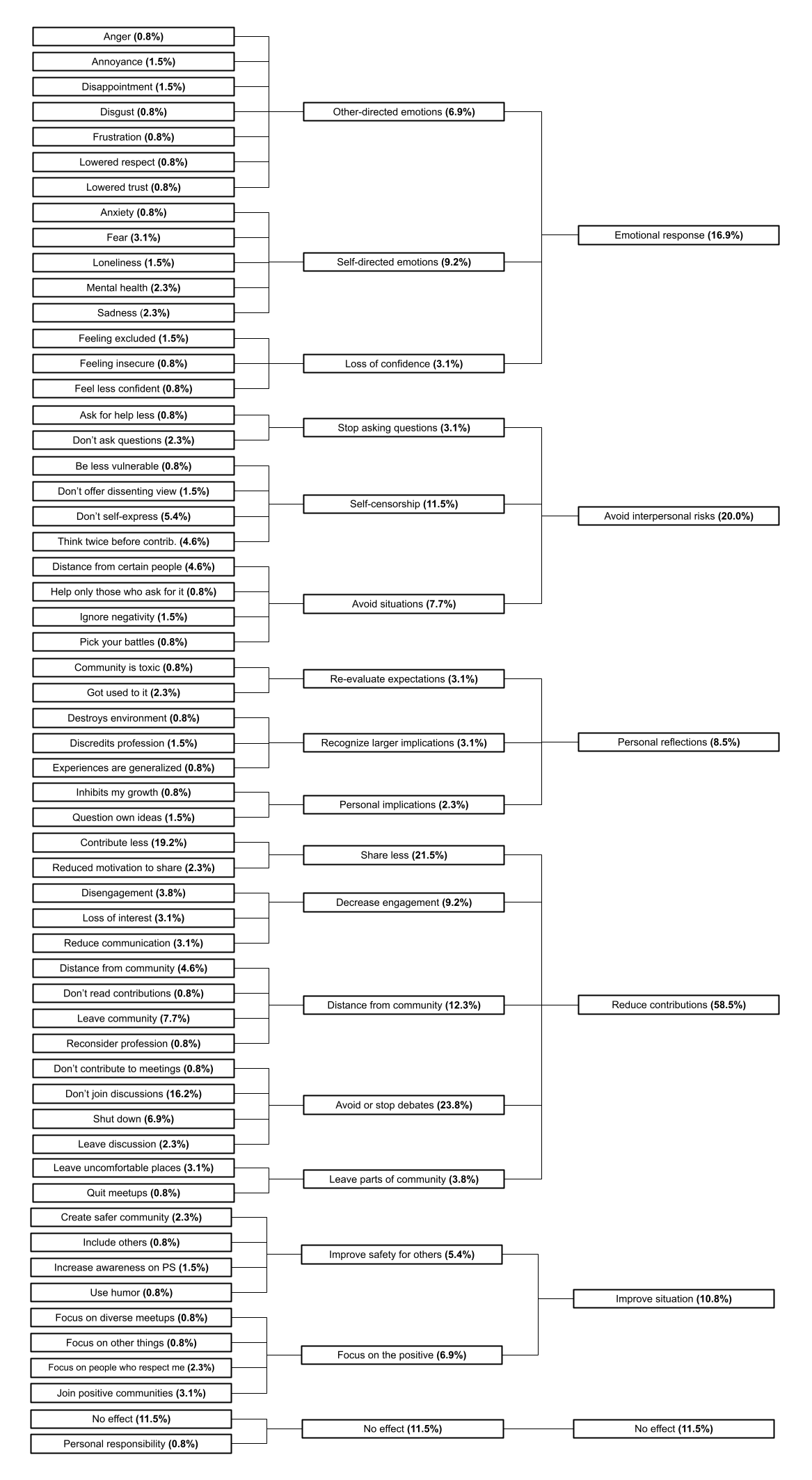}
\Description[Gioia schema for the effects of low psychological safety]{The key insights from this picture are described in this section.}
\label{fig:gioia-effects}
\end{figure}

\begin{figure}[H]
\centering
\caption{First-order concepts, second-order themes, and aggregate dimensions in the behaviors that contributed to lowered psychological safety as reported in Communities of Practice by participants ($N=120$)}
\includegraphics[width=3.5in]{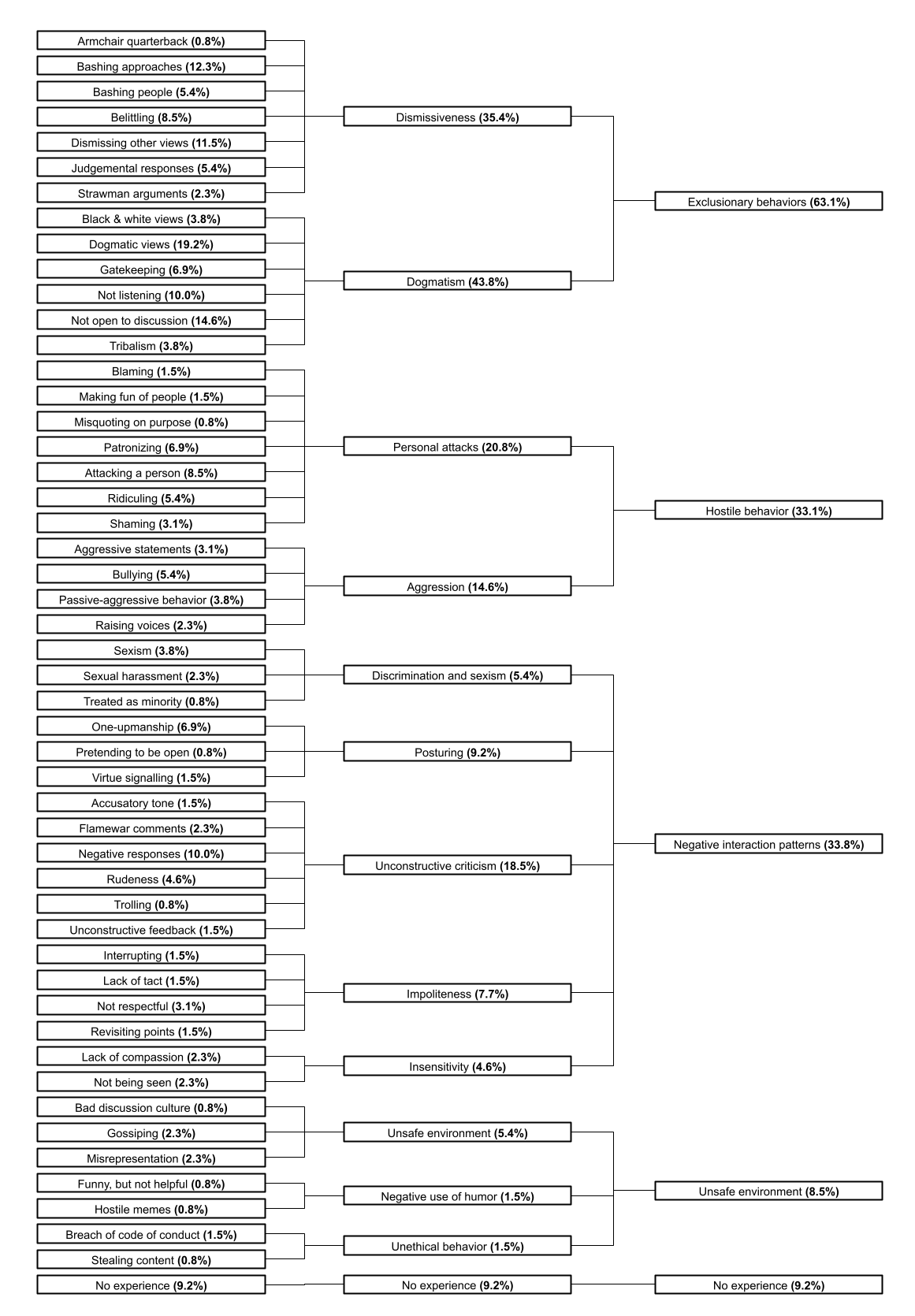}
\Description[Gioia schema for behaviors that decrease psychological safety]{The key insights from this picture are described in this section.}
\label{fig:gioia-experience}
\end{figure}

\begin{figure}[H]
\centering
\caption{First-order concepts, second-order themes, and aggregate dimensions in the factors contributing to lower psychological safety in Communities of Practice as reported by participants ($N=118$)}
\includegraphics[width=3.5in]{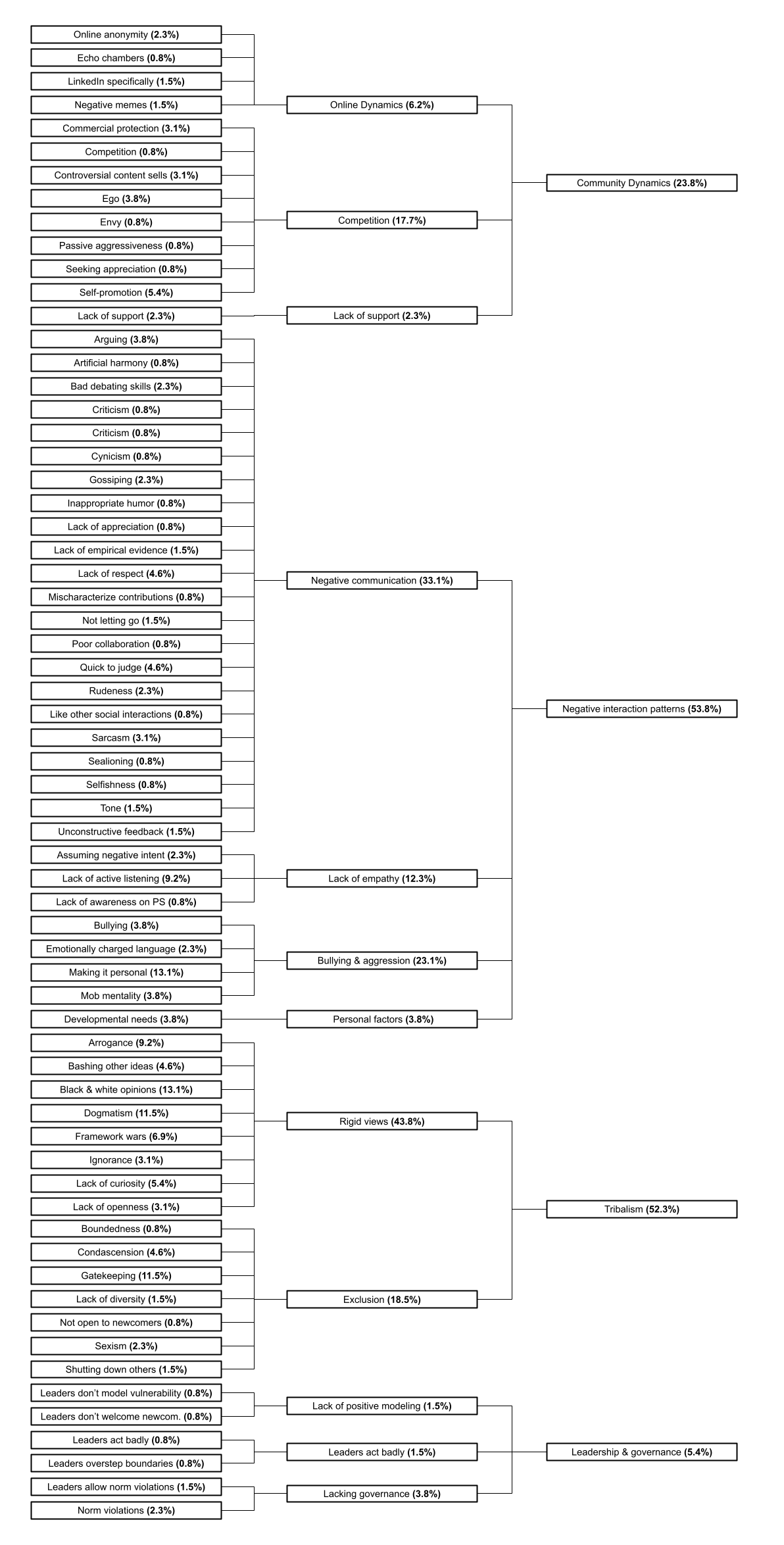}
\label{fig:gioia-factors}
\Description[Gioia schema for factors that decrease psychological safety]{The key insights from this picture are described in this section.}
\end{figure}

\begin{figure}[H]
\centering
\caption{First-order concepts, second-order themes, and aggregate dimensions in the strategies to improve psychological safety in Communities of Practice as suggested by participants ($N=111$)}
\includegraphics[width=3.5in]{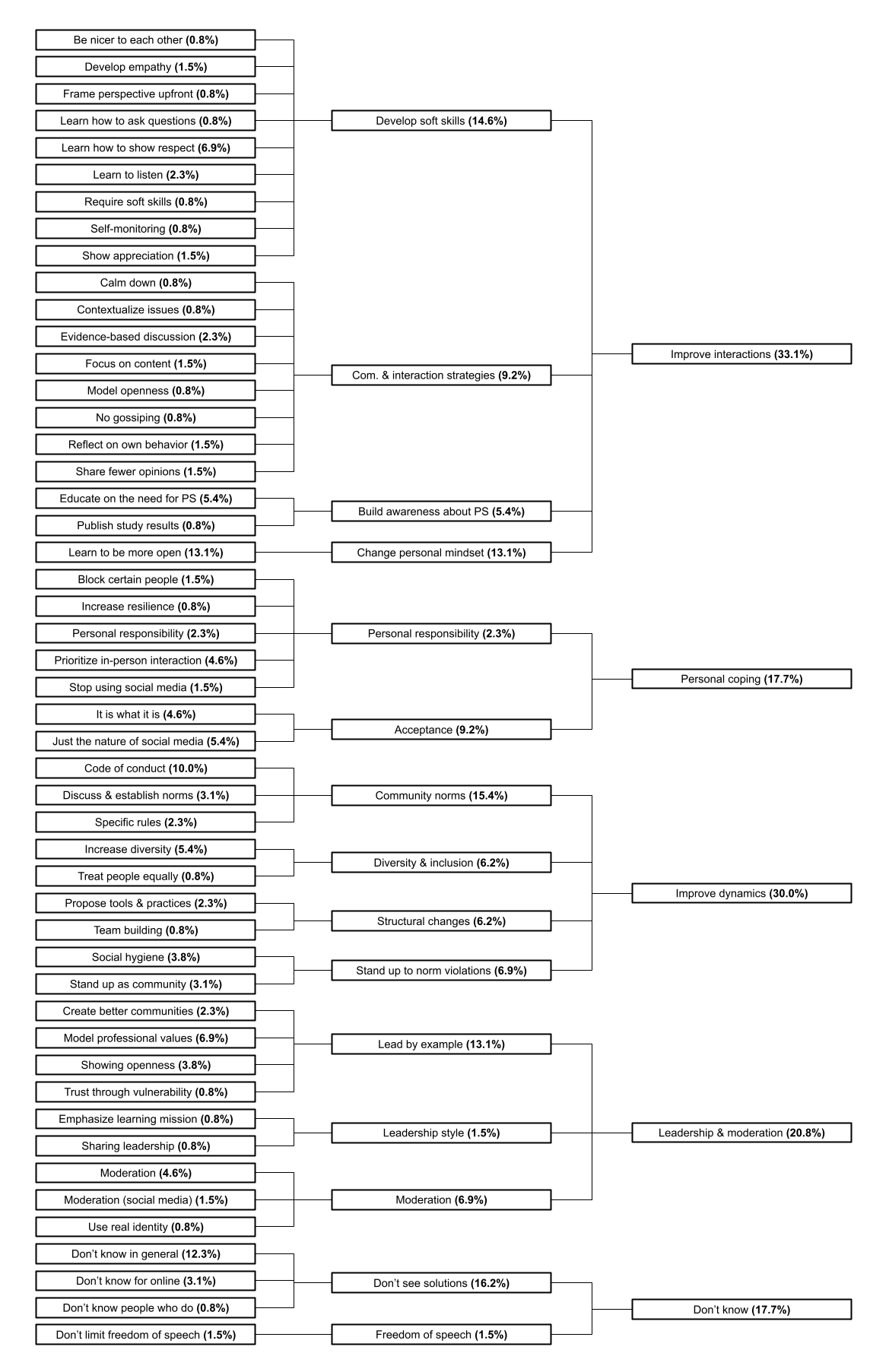}
\label{fig:gioia-strategies}
\Description[Gioia schema for strategies to improve psychological safety]{The key insights from this picture are described in this section.}
\end{figure}

\begin{table}[H]
\tiny
\label{tab:appendix-phase1-descriptives}
\centering
\caption{N, Means, Standard Deviations, Skewness, Kurtosis and Correlations (Pearson) for continuous and ordinal variables in Phase~I. Correlations marked with an asterisk (*) are statistically significant at $p <.01$.}
\begin{tabular}{lccccccccc}
\toprule
\textbf{Variable} & \textbf{N} & \textbf{Mean} & \textbf{SD} & \textbf{Skewness} & \textbf{Kurtosis} & \textbf{1} & \textbf{2} & \textbf{3} & \textbf{4} \\
\midrule
1. Psychological safety (face-to-face) & 141 & 3.88 & .88 & -1.03 & 1.14 & 1.00 &  &  &  \\
2. Psychological safety (online) & 143 & 2.93 & 1.03 & -.13 & -.77 & .45* & 1.00 &  &  \\
\multicolumn{10}{c}{\textit{Ordinal demographic variables}} \\
3. Age group & 143 & 3.82 & 1.13 & .69 & .72 & -.04 & .08 & 1.00 &  \\
4. Seniority in community & 143 & 4.35 & .74 & -.98 & .57 & -.09 & -.07 & .38* & 1.00 \\
\bottomrule
\end{tabular}
\end{table}

\begin{table}[!ht]
\centering
\small
\caption{Results of Confirmatory Factor Analysis from individual-level responses ($n=143$). Principal Components Analysis with Oblimin rotation and Kaiser normalization. Items marked with asterisk (*) were removed in the final scale. }
\label{tab:appendix-phase1-cfa}
\begin{tabular}{@{}m{6cm}m{1.5cm}m{1.5cm}@{}}
\toprule
\textbf{Item} & \textbf{1} & \textbf{2} \\
\midrule
Psychological safety 1 (Face-to-face) & .905 &   \\ \addlinespace
Psychological safety 2 (Face-to-face) & .886 &   \\ \addlinespace
Psychological safety 3 (Face-to-face) & .930 &   \\ \addlinespace
Psychological safety 4 (Face-to-face) & .592 &   \\ \addlinespace
Psychological safety 5 (Face-to-face)* & .356 &  \\ \addlinespace
Psychological safety 6 (Face-to-face) & .837 \\ \addlinespace
Psychological safety 1 (Online) & & .755\\ \addlinespace
Psychological safety 2 (Online) & & .814 \\ \addlinespace
Psychological safety 3 (Online) & & .869 \\ \addlinespace
Psychological safety 4 (Online) & & .606 \\ \addlinespace
Psychological safety 5 (Online)* & & .370 \\ \addlinespace
Psychological safety 6 (Online) & & .877 \\ \addlinespace
\bottomrule
\end{tabular}
\vspace{2em}
\end{table}

\begin{table}[]
\tiny
\caption{Phase~I Questionnaire}
\label{tab:appendix-phase1-survey}
\begin{tabularx}{\textwidth}{@{}p{3cm}p{4.5cm}p{1cm}p{4.3cm}@{}}
\toprule
\textbf{Variable} & \textbf{Question} & \textbf{Type} & \textbf{Options} \\
\midrule
\multicolumn{4}{l}{\textit{How frequently do you share something with the Agile Community on the following social media? (e.g. post a comment, article, blogpost, video):}} \\
Where do you meet (online) & LinkedIn & Categorical & Never, Less than once a month, At least once a month, At least once a   week, At least once per day \\
Where do you meet (online) & Twitter / X & Categorical & (as above) \\
Where do you meet (online) & Medium & Categorical &  (as above) \\
Where do you meet (online) & Mastodon & Categorical &  (as above) \\
Where do you meet (online) & Reddit & Categorical &  (as above) \\
Where do you meet (online) & Discord \& Slack & Categorical &  (as above) \\
Where do you meet (online) & Agile forums & Categorical & (as above) \\
Where do you meet (online) & YouTube & Categorical &  (as above) \\
\multicolumn{4}{l}{\textit{On the social medium you use most frequently, how do you experience the following?}} \\
Psychological safety online \#1 & On this medium, members of the Agile community foster an   environment where it easy to ask for help & Likert (1-5) & Strongly disagree, Somewhat disagree, Neither agree nor disagree,   Somewhat agree, Strongly agree \\
Psychological safety online \#2 & On this medium, members of the Agile community are open to   different perspectives and opinions & Likert (1-5) &  (as above) \\
Psychological safety online \#3 & On this medium, people in the Agile community make an effort   to understand each others' views & Likert (1-5) &  (as above) \\
Psychological safety online \#4 & On this medium, I frequently observe or experience behavior in   the Agile community that lowers psychological safety & Likert (1-5) & (as above, reversed) \\
Psychological safety online \#5 & On this medium, it is important to me that interactions in the   Agile community are psychologically safe & Likert (1-5) &  (as above) \\
Psychological safety online \#6 & On this medium, I experience psychological safety in the Agile community to be high & Likert (1-5) &  (as above) \\
\multicolumn{4}{l}{\textit{Where do you meet members of the Agile community face-to-face (video or in-person)?}} \\
Where do you meet (face-to-face) & Conferences & Categorical & Never, Rarely, Often, Very Often \\
Where do you meet (face-to-face) & Online meetups & Categorical & (as above) \\
Where do you meet (face-to-face) & In-person meetups & Categorical & (as above) \\
Where do you meet (face-to-face) & In workshop or training & Categorical & (as above) \\
Where do you meet (face-to-face) & Other & Categorical & (as above) \\
\multicolumn{4}{l}{\textit{During face-to-face interactions with members of the Agile community, how do you experience the following?}} \\
Psychological safety face-to-face \#1 & During face-to-face interactions, members of the Agile   community foster an environment where it easy to ask for help & Likert (1-5) & Strongly disagree, Somewhat disagree, Neither agree nor disagree,   Somewhat agree, Strongly agree \\
Psychological safety face-to-face \#2 & During face-to-face interactions, members of the Agile   community are open to different perspectives and opinions & Likert (1-5) &  (as above) \\
Psychological safety face-to-face \#3 & During face-to-face interactions, people in the Agile   community make an effort to understand each others' views & Likert (1-5) &  (as above) \\
Psychological safety face-to-face \#4  & During face-to-face interactions, I frequently observe or experience behavior in the Agile community that lowers psychological safety & Likert (1-5) & (as above, reversed) \\
Psychological safety face-to-face \#5 & During face-to-face interactions, it is important to me that   interactions in the Agile community are psychologically safe & Likert (1-5) & (as above) \\
Psychological safety face-to-face \#6 & During face-to-face interactions, I experience psychological   safety in the Agile community to be high & Likert (1-5) & (as above) \\
Experience with low psychological safety & What are examples of behavior you have personally encountered during interactions in the Agile community that lowered psychological safety for you? & Essay & n/a \\
Factors that decrease psychological safety & In your experience, what behavior or factors decrease psychological safety in the Agile community? & Essay & n/a \\
Effects of low psychological safety & How does a lack of psychological safety in interactions in the   Agile community affect you or your behavior? & Essay & n/a \\
Strategies to improve psychological safety & What measures or strategies could improve psychological safety and encourage more respectful and constructive discussions within the Agile community? & Essay & n/a \\
Role & What best describes your role? & Categorical & Developer, analyst or tester or designer, Scrum Master, Agile Coach,Trainer or facilitator, Consultant, Other, Prefer not to say \\
Age group & What is your age group & Categorical & 18-25 years, 26-35 years, 36-45 years, 46-55 years, 56-65 years, 66+ years, Prefer not to say \\
Seniority & How many years have you been active in the Agile community? & Categorical & Less then a year, Between 1 and 2 years, Between 2 and 5 years, Between 5   and 10 years, More than 10 years \\
Gender & What is your gender? & Categorical & Male, female, non-binary or third-gender, prefer not to say \\
Content creatorship & Do you create content for the Agile community (videos, blogs,   products, etc?) & Categorical & Yes, No \\
\bottomrule
\end{tabularx}
\end{table}

\begin{table}[]
\tiny
\caption{Phase~II Questionnaire}
\label{tab:appendix-phase2-survey}
\begin{tabularx}{\textwidth}{@{}p{3cm}p{4.5cm}p{1cm}p{4.3cm}@{}}
\toprule
\textbf{Variable} & \textbf{Question} & \textbf{Type} & \textbf{Options} \\
\midrule
Psychological safety face-to-face vs   online & How do these results align with your expectations? & Likert (1-5) & The results strongly differ from my expectations, The results somewhat   differ from my expectations, The results neither differ from nor match my   expectations, The results somewhat match my expectations, The results   strongly match my expectations \\
 & What is your interpretation of this result? & Essay & n/a \\
Psychological safety by gender & How do these results align with your expectations? & Likert (1-5) & The results strongly differ from my expectations, The results somewhat   differ from my expectations, The results neither differ from nor match my   expectations, The results somewhat match my expectations, The results   strongly match my expectations \\
 & What is your interpretation of this result? & Essay & n/a \\
Psychological safety by role & How do these results align with your expectations? & Likert (1-5) & The results strongly differ from my expectations, The results somewhat   differ from my expectations, The results neither differ from nor match my   expectations, The results somewhat match my expectations, The results   strongly match my expectations \\
 & What is your interpretation of this result? & Essay & n/a \\
Psychological safety by age & How do these results align with your expectations? & Likert (1-5) & The results strongly differ from my expectations, The results somewhat   differ from my expectations, The results neither differ from nor match my   expectations, The results somewhat match my expectations, The results   strongly match my expectations \\
 & What is your interpretation of this result? & Essay & n/a \\
Follow-up questions & Based on the results you've interpreted so far, what follow-up questions   come to mind for you? & Essay & n/a \\
Prior participation in Phase~I & Did you participate in the survey on psychological safety yourself? & Categorical & Yes, No, I do not remember \\
Interest in research process & Did you find this process interesting? & Likert (1-5) & Not interesting at all, Slightly interesting, Moderately interesting,   Very interesting, Extremely interesting \\
\bottomrule
\end{tabularx}
\end{table}

\end{appendix}

\end{document}